%% file: main.tex
\documentclass[
 reprint,
showpacs,
preprintnumbers,
 amsmath,amssymb,
 aps,
 superscriptaddress,
prd]{revtex4-1}

\usepackage{graphicx}
\graphicspath{{figures/}}
\usepackage{dcolumn}
\usepackage{bm}
\usepackage{parskip}
\usepackage{multirow}
\usepackage{color}
\usepackage[normalem]{ulem}

\usepackage[compact]{titlesec}
\titlespacing*{\section}{0pt}{1.1\baselineskip}{\baselineskip}
\titlespacing*{\subsection}{0pt}{1.1\baselineskip}{\baselineskip}

\DeclareRobustCommand{\rchi}{{\mathpalette\irchi\relax}}
\newcommand{\irchi}[2]{\raisebox{0.95\depth}{$#1\chi$}}
\usepackage{enumitem}

\begin{document}
\setlength{\abovedisplayskip}{6pt}
\setlength{\belowdisplayskip}{6pt}
\setlength{\abovedisplayshortskip}{6pt}
\setlength{\belowdisplayshortskip}{6pt}
\setlength{\parskip}{0pt}
\setlength{\parindent}{18pt}
\newcommand{\keVee}{keV$_{\textrm{ee}}$}
\newcommand{\keVnr}{keV$_{\textrm{nr}}$}

\title{An Effective Field Theory Analysis of the First LUX Dark Matter Search}
\include{lux-tex-author-list_include}

\collaboration{The LUX Collaboration}

\date{\today}

\begin{abstract}
\noindent The Large Underground Xenon (LUX) dark matter search was a 250-kg active mass dual-phase time projection chamber that operated by detecting light and ionization signals from particles incident on a xenon target.  In December 2015, LUX reported a minimum 90\% upper C.L.~of $6\times10^{-46}$ cm${^2}$ on the spin-independent WIMP-nucleon elastic scattering cross section based on a $1.4\times10^{4}$ kg$\cdot$day exposure in its first science run.  Tension between experiments and the absence of a definitive positive detection suggest it would be prudent to search for WIMPs outside the standard spin-independent/spin-dependent paradigm.  Recent theoretical work has identified a complete basis of 14 independent effective field theory (EFT) operators to describe WIMP-nucleon interactions.  In addition to spin-independent and spin-dependent nuclear responses, these operators can produce novel responses such as angular-momentum-dependent and spin-orbit couplings.  Here we report on a search for all 14 of these EFT couplings with data from LUX's first science run. Limits are placed on each coupling as a function of WIMP mass. 
\end{abstract}

\maketitle

\section{\label{section:intro}Introduction}

The existence of a non-luminous, non-baryonic matter component to the universe is supported by a wealth of astrophysical data ranging from galactic to cosmological scales\,\cite{rubin1970, clowe2006, SDSS2014, PLANCK2015overview, hu2002}.  Under the $\Lambda$CDM model, dark matter forms the majority (84.3\%) of the matter density and a substantial portion (25.9\%) of the energy density of the universe \cite{PLANCK2015}.  Although some of its properties have been constrained through astrophysical measurements, its exact nature remains one of the most intriguing mysteries of modern physics.  A favored dark matter candidate is the Weakly Interacting Massive Particle (WIMP), which may have been produced thermally in the early universe and arises naturally in supersymmetric and Kaluza-Klein theories.  The direct detection of WIMPs through their scattering off of nuclei is one of the most promising avenues for detection \cite{goodmanwitten1985, feng2010}.  WIMPs are expected to have a mass in range of $\mathcal{O}$(GeV) to $\mathcal{O}$(TeV) and are therefore kinematically well-matched to atomic nuclei.  WIMP-nucleus recoils involve energy transfers on the order of 10 keV.  Although rare, such energy depositions may be detectable in a laboratory target.

The Large Underground Xenon (LUX) experiment was a 250-kg active mass dual-phase (liquid/gas) xenon-based time projection chamber deployed 4850 feet underground (4300 meters water equivalent) at the Sanford Underground Research Facility in Lead, South Dakota.  Particle interactions in xenon produce an abundant signal in both scintillation and ionization channels. Xenon is transparent to its own scintillation light, so interactions can be detected with little loss of sensitivity even at the center of very large volumes.  Similarly, a high ionization yield and long electron drift lengths in xenon enable charge signals to be efficiently extracted.  Xenon is also an intrinsically low-background material.  Of its naturally-occurring radioactive isotopes, $^{127}$Xe has a half-life of only 36 days, and all but one other decay more rapidly.  After a few weeks, these backgrounds therefore become insignificant.  $^{136}$Xe undergoes double-beta decay with a half-life of $2.11\times10^{21}$ years\,\cite{EXO2011} and constitutes a negligible background for all current dark matter searches.  Finally, its high atomic number ($Z=54$), atomic mass ($A\sim131$), and high natural abundance of $n$-odd isotopes $^{129}$Xe ($26.4\%$) and $^{131}$Xe ($21.2\%$) give high sensitivity to both spin-independent (SI) and spin-dependent (SD) WIMP-nucleon interactions.  Leveraging these advantages as well as a number of hardware innovations and novel calibration methods, LUX set world-leading constraints on SI WIMP-nucleon and SD WIMP-neutron interaction cross sections for a wide range of WIMP masses\,\cite{luxrun3re2016, luxrun42016, luxsd2016, LUX_SD_Run3_Run4}.

Most other direct dark matter searches have likewise detected no evidence of WIMP-nucleus interactions to date.  In contrast, the DAMA/LIBRA experiment has reported a $12.9\sigma$ event excess consistent with an annual modulation signal from a light WIMP\,\cite{Bernabei:2018yyw}.  Many attempts have been made to reconcile this event excess with LUX and other null results through the application of increasingly more exotic dark matter interaction models beyond the typical experimental treatment of WIMP-nucleon elastic scattering, which focuses only on interactions governed by SI and SD operators.  Examples include ``xenophobic'' isospin-violating dark matter \cite{kumar2013}, and form factor dark matter, which has interactions modulated by momentum-dependent form factors \cite{feldstein2010}.  However, this problem may be treated in a more general way, avoiding model-specific assumptions about the properties of dark matter and ensuring full parameter space coverage.

Traditional WIMP direct-detection phenomenology retains only the WIMP-nucleon SI and SD interaction terms, as these interactions do not depend on the momentum transfer of the interaction and therefore feature a finite cross-section in the limit of zero momentum transfer. However, this limit can be inappropriate for two reasons: (1) the Fermi motion of nucleons inside nuclei renders the static limit inaccurate \cite{hax1}; (2) for larger energy nuclear recoils, the momentum transferred to the nucleus can be significant.  As a result, the standard SI and SD interactions may be subject to corrections from momentum- or velocity-dependent operators.  In addition, the standard operators may also be suppressed in such a way that the dominant interaction is momentum- or velocity-dependent, as in a composite dark matter scenario where the WIMP-nucleon interactions are governed primarily by dipole or anapole operators \cite{greshamzurek2014}.

In this paper, we summarize recent theoretical work by Fitzpatrick et al.~that that addresses the shortcomings of the assumption of purely static nucleons and approaches the problem of WIMP-nucleon elastic scattering in a model-independent way. We use the first LUX data set to constrain the coupling strengths of all terms in a complete basis which spans all possible forms of the WIMP-nucleon interactions \cite{hax1,hax2,hax3,hax4}.  This is a much richer set of interactions than those included in the standard SI/SD paradigm, in particular yielding a set of entirely new nuclear responses with different strengths in different target nuclei.  We constrain the coupling strengths of dark matter to nucleons through each of these operators. 

\section{\label{section:EFTframework}An Effective Field Theory Framework for Direct Detection}

In general, the differential event rate with respect to recoil energy for WIMP-nucleon interactions is given by \cite{kamionkowski1996}:
\begin{align} \label{eqn:dRdEReftch}
\frac{dR}{dE_R} = \frac{\rho_0}{m_{\rchi} m_A} \int_{v>v_{min}} v f(\vec{v}) \frac{d\sigma}{d E_R} d^3v 
\end{align}
where $\rho_0$ is the local density of WIMPs in the galactic halo, $m_{\rchi}$ and $m_A$ are the masses of the WIMP and target nucleus respectively, $d\sigma/dE_R$ is the differential WIMP-nucleus interaction cross section with respect to recoil energy, $\vec{v}$ is the velocity of the incident WIMP with respect to the target, $f(\vec{v})$ describes the WIMP velocity distribution, and $v_{min}$ is the minimum WIMP velocity needed to create a recoil of energy $E_R$.  The details of the underlying particle physics are contained in the differential cross section $\frac{d\sigma}{dE_R}$:
\begin{align} \label{eqn:fermisgoldenrule_eftch2}
\frac{d\sigma}{dE_R} = & \frac{1}{32 \pi v^2} \frac{1}{m_{\rchi}^2 m_A} \frac{1}{(2j_A+1)(2j_{\rchi}+1)} \times & \sum_{\textrm{Spins}} |\mathcal{M}|^2
\end{align}
where $\mathcal{M}$ is the WIMP-nucleus scattering amplitude, and we average over the initial nuclear spin $j_A$ and WIMP spin $j_{\rchi}$ and sum over final spins.  A factor of $1/(4 m_{\rchi} m_A)^2$ is introduced to account for the normalization used in matching relativistic WIMP-nucleon interaction operators to the corresponding nonrelativistic operators \cite{hax1}.

In \cite{hax1,hax2,hax3,hax4}, the WIMP scattering amplitude $|\mathcal{M}|^2$ is calculated by modeling each scatter as a four-particle contact interaction.  The interaction Lagrangian has the generic form
\begin{align} \label{eqn:eftlagrangian}
	\mathcal{L}_{\textrm{int}} = \bar{\rchi} \mathcal{O}_{\rchi} \rchi \bar{N} \mathcal{O}_{N} N \equiv \mathcal{O} \bar{\rchi} \rchi \bar{N} N
\end{align}
where $\rchi$ and $N$ are nonrelativistic fields denoting the incident WIMP and the target nucleon, respectively.  Although we do not consider WIMP inelastic scattering, note that it can be treated by generalizing to $\bar{\rchi}_1 \mathcal{O}_{\rchi} \rchi_2$, where  ${\rchi}_1$ and ${\rchi}_2$ have different masses.  

Under conservation of momentum and Galilean invariance, the four momenta of the particles can be reduced to a basis of two independent quanitites, chosen for convenience to be the Hermitian quantities $i\vec{q}$, where $\vec{q}$ is the momentum transfer imparted from the incident WIMP to the target nucleon, and $\vec{v}^{\perp} = \vec{v} + \vec{q}/2\mu_N$, where $\mu_N = m_\rchi m_N (m_\rchi + m_N)^{-1}$ is the WIMP-nucleon reduced mass.  $\vec{v}^{\perp}$ is the component of WIMP incident velocity $\vec{v}$ transverse to $\vec{q}$.  All WIMP-nucleon operators subject to these basic symmetries can be written as a combination of $i\vec{q}$, $\vec{v}^{\perp}$, the nucleon spin $\vec{S}_N$, and the WIMP spin $\vec{S}_{\rchi}$.  For a WIMP-nucleon interaction that involves the exchange of a spin-0 or spin-1 mediator, this yields 11 possible combinations:
\begin{align}
	\mathcal{O}_{1} & = 1 \nonumber \\
	\mathcal{O}_{2} & = (v^{\perp})^2 \nonumber \\
	\mathcal{O}_{3} & = i \vec{S}_N \cdot (\vec{q} \times \vec{v}^{\perp}) \nonumber \\
	\mathcal{O}_{4} & = \vec{S}_{\rchi} \cdot \vec{S}_N \nonumber \\
	\mathcal{O}_{5} & = i \vec{S}_{\rchi} \cdot (\vec{q} \times \vec{v}^{\perp}) \nonumber \\
	\mathcal{O}_{6} & = (\vec{S}_{\rchi} \cdot \vec{q}) (\vec{S}_{N} \cdot \vec{q}) \nonumber \\
	\mathcal{O}_{7} & = \vec{S}_N \cdot \vec{v}^{\perp} \nonumber \\
	\mathcal{O}_{8} & = \vec{S}_{\rchi} \cdot \vec{v}^{\perp} \nonumber \\
	\mathcal{O}_{9} & = i \vec{S}_{\rchi} \cdot (\vec{S}_N \times \vec{q}) \nonumber \\
	\mathcal{O}_{10} & = i \vec{S}_N \cdot \vec{q} \nonumber \\
	\mathcal{O}_{11} & = i \vec{S}_{\rchi} \cdot \vec{q} 
\end{align}
There are five additional exotic operators that arise only in interactions not involving the exchange of a spin-0 or spin-1 mediator:
\begin{align}
	\mathcal{O}_{12} & = \vec{S}_{\rchi} \cdot (\vec{S}_N \times \vec{v}^{\perp}) \nonumber \\
	\mathcal{O}_{13} & = i(\vec{S}_{\rchi} \cdot \vec{v}^{\perp})(\vec{S}_{N} \cdot \vec{q}) \nonumber \\
	\mathcal{O}_{14} & = i(\vec{S}_{\rchi} \cdot \vec{q})(\vec{S}_{N} \cdot \vec{v}^{\perp}) \nonumber \\
	\mathcal{O}_{15} & = -(\vec{S}_{\rchi} \cdot \vec{q})((\vec{S}_{N} \times \vec{v}^{\perp}) \cdot \vec{q}) \nonumber \\
	\mathcal{O}_{16} & = -((\vec{S}_{\rchi} \times \vec{v}^{\perp}) \cdot \vec{q})(\vec{S}_{N} \cdot \vec{q})
\end{align}
Operator $\mathcal{O}_2$ to leading order does not arise in the nonrelativistic limit to any relativistic operator, and operator $\mathcal{O}_{16}$ can be written as a linear combination of operators $\mathcal{O}_{12}$ and $\mathcal{O}_{15}$.  A generic WIMP-nucleon interaction $\mathcal{L}_{\textrm{int}} = \sum_i c_i \mathcal{O}_i$ is written as a sum over the fourteen basis operators $\mathcal{O}_{1}$ and $\mathcal{O}_{3},\ldots \mathcal{O}_{15}$, with $c_i$ denoting the coupling constant associated with operator $\mathcal{O}_i$.

These 14 operators acting on an individual nucleon's available degrees of freedom $\vec{S}_N$ and $\vec{v}^{\perp}_N$ can produce six distinct nuclear charges and currents.
\begin{align} \label{eqn:nuclchargesandcurrents}
1, \quad \vec{v}^{\perp}_N \cdot \vec{v}^{\perp}_N, \quad \vec{S}_N \cdot \vec{v}^{\perp}_N, \\ \nonumber
\vec{S}_N, \quad \vec{v}^{\perp}_N, \quad \textrm{and } \vec{S}_N \times \vec{v}^{\perp}_N
\end{align}
Here $\vec{v}^{\perp}_N \cdot \vec{v}^{\perp}_N$ can be neglected to lowest order.  For elastic scatters, nuclear selection rules for parity and time reversal constrain most off-diagonal terms in the scattering amplitude to be zero.  Creating a Lagrangian formed from the operators listed in Eq.~4  and Eq.~5, and then grouping the terms in terms of the nuclear charges and currents of Eq.~6, leads to six independent nuclear responses contributing to the total scattering amplitude.
Each response is proportional to one of the nuclear charges or currents and has an interpretation familiar from standard electroweak physics or simple extensions thereof \cite{hax1,hax2}.  We denote the six possible nuclear responses as M, $\Sigma'$, $\Sigma''$, $\Delta$, $\widetilde{\Phi}'$, and $\Phi''$, and in the leading-multipole, long-wavelength limit they behave as follows:
\\
\begin{description}[leftmargin=0.0\parindent, labelindent=0.0\parindent, before={\renewcommand\makelabel[1]{##1}}, after=\newline]
	\item [$\textbf{M}$] is a spin-independent nuclear response proportional to $Z$ (for protons) or $(A-Z)$ (for neutrons). It arises, for example, from the scalar interaction $\mathcal{O}_1$, and approaches $\frac{A}{2\sqrt{\pi}}$ in the long-wavelength limit.  The coupling coefficient $c_1$ is related to the standard SI WIMP-nucleon zero-momentum-transfer cross-section $\sigma_{0,\textrm{SI}} $ by:
	\begin{align}
	\sigma_{0,\textrm{SI}} =  \left( \frac{4\mu_A^2}{\pi} \right) [c_1^p Z + c_1^n (A-Z)]^2
	\end{align}
Other interactions (for example, $\mathcal{O}_{11}$, which is momentum-dependent and therefore neglected in the standard SI/SD treatment) can also give rise to a spin-independent nuclear response. Of all direct detection targets, heavy targets like iodine and xenon are the most sensitive to the M response. \\
	\item [\boldmath$\Sigma''$ and $\Sigma'$] depend on the component of the nuclear spin longitudinal and transverse to the momentum transfer ($\vec{q}$), respectively.  In the long-wavelength limit, they approach $ \frac{1}{2\sqrt{3\pi}} \sum_{i=1}^A \sigma(i) \textrm{ and } \sum_{i=1}^A \sigma(i)$, respectively, where $\sigma(i)$ denotes the spin operator acting on the $i$th nucleon in the target nucleus, and are proportional to the expected proton or neutron spin content of the nucleus,  $\langle S_p \rangle$ or $\langle S_n \rangle$.  As $q^2$ becomes non-negligible (departing from the long-wavelength limit) their form factors differ significantly, and several of the WIMP-nucleon interaction operators (for example, $\mathcal{O}_9$ and $\mathcal{O}_{10}$) give rise to one but not the other.  A particular linear combination ($\Sigma'' + \Sigma'$) arises from the standard SD interaction $\mathcal{O}_4 = \vec{S}_{\rchi} \cdot \vec{S}_N$.  For spin-$1/2$ dark matter, the spin-dependent coupling coefficient $c_4$ can be written in terms of the standard SD WIMP-nucleon coupling strengths $a_p$, $a_n$ as follows \cite{hax1}:
	\begin{align}
		c_4^N = 32\sqrt{2} m_N m_{\rchi} G_F a_N \textrm{ for } N=n,p
	\end{align}
$c_4^N$ is related to the standard SD zero-momentum-transfer interaction cross section $\sigma_{0,\textrm{SD}}$ by the formula:
	\begin{align}
		\sigma_{0,\textrm{SD}} = \frac{3}{256 \pi} \frac{\mu_N^2}{m_N^2 m_{\rchi}^2} c_4^{N2} \textrm{  for } N=n \textrm{ or } p
	\end{align}
Nuclei with unpaired protons (for example, $^{19}$F) or unpaired neutrons (for example, $^{129}$Xe or $^{131}$Xe) are the most sensitive to $\Sigma''$ and $\Sigma'$.
\end{description} 

\noindent {\boldmath$\Delta$}, {\boldmath$\Phi''$}, and {\boldmath$\widetilde{\Phi}'$} are novel responses.  Each contains explicit factors of $q^2$ and arises from the velocities of nucleons $\vec{v}_{N,i}^{\perp}$ inside the nucleus.  Consequently, these new responses do not appear in the point-nucleus limit or the standard SI/SD treatment at all.  In more detail:
\\
\begin{description} [leftmargin=0.0\parindent, labelindent=0.0\parindent, before={\renewcommand\makelabel[1]{##1}}, after=\newline]
	\item[\boldmath$\Delta$] depends on the angular-momentum content of the nucleus, $\langle L_p\rangle$ or $\langle L_n\rangle$, which in turn depend on the relative nucleon velocities.  $\Delta$ approaches $\frac{1}{2\sqrt{6\pi}}\sum_{i=1}^Al(i)$ in the long-wavelength limit, where $l(i)$ denotes the angular momentum operator acting on the $i$th nucleon in the target nucleus, and therefore gains a kinematic enhancement of $A$, being most favorable to heavy targets with an unpaired nucleon in a non-$s$-shell orbital.\\
	\item[\boldmath $\Phi''$] depends on the spin-orbit coupling between a nucleons spin and its own angular momentum.  It approaches $\frac{1}{6\sqrt{\pi}}\sum_{i=1}^A[\vec{\sigma}(i)\cdot\vec{l}(i)]$ in the long-wavelength limit, which is proportional to $\langle\vec{S}_p\cdot\vec{L}_p \rangle$ or $\langle\vec{S}_n\cdot\vec{L}_n \rangle$. $\Phi''$ has a kinematic enhancement of $A$ and favors targets with unfilled angular momentum orbitals.\\
	\item[\boldmath$\widetilde{\Phi}'$] is a tensor operator that is proportional to $\Phi''$ in the long-wavelength limit.  It appears only in models with the operators $\mathcal{O}_{12}$ and $\mathcal{O}_{13}$, which cannot arise from the exchange of a spin-0 or spin-1 mediator.  $\widetilde{\Phi}'$ only contributes in targets with $j_A \geq 1$.  This includes $^{131}$Xe, which has spin 3/2.
\end{description}
	
\noindent Finally, two pairs of nuclear responses (M and $\Phi''$) and ($\Sigma'$ and $\Delta$) can interfere with each other.  These interference responses arise only in off-diagonal terms in the scattering amplitude matrix. Only pairs of operator $\mathcal{O}_i$ and $\mathcal{O}_j$ that have the same parity and time-reversal properties can interfere.  These pairs are $(i,j)$ = $(1,3)$, $(4,5)$, $(4,6)$, $(8,9)$, $(11,12)$, $(11,15)$, and $(12,15)$.  All other off-diagonal terms vanish.  

A WIMP-nucleon interaction Lagrangian $\mathcal{L}_{\textrm{int}}^{(N)} = \sum_i c_i^{(N)} \mathcal{O}_i^{(N)}$, with $N$ = $n$ or $p$  denoting the type of nucleon involved in the interaction, gives rise to the following total WIMP-nucleus scattering amplitude.

\begin{align} \label{eqn:defformfactors}
	\frac{1}{(2j_A+1)(2j_{\rchi}+1)} & \sum_{\textrm{Spins}} |\mathcal{M}|^2 \nonumber \\
	\equiv \frac{m_A^2}{m_N^2} \times \sum_{i,j} & \sum_{(N,N')} c_i^{N} c_j^{N'} F_{i,j}^{(N,N')}
\end{align}
where $j_A$ is the spin content of the target nucleus, $j_\rchi$ is the WIMP spin, $c_i$ is the coefficient of operator $\mathcal{O}_i$ in the Lagrangian, and the effective field theory form factors $F_{i,j}^{(N,N')} = F_{i,j}^{(N,N')}(v^2, q^2)$ are defined to be the coefficient of $c_i c_j$.  $F_{i,j}$ can depend on velocity $v$ and momentum transfer $q$ and is different for different target nuclei \cite{hax1}.  The explicit sum over nucleon pairs $(N, N') = (n,n),(n,p),(p,n),(p,p)$ takes into account two-body currents in the nucleus.

Each form factor $F_{i,j}^{(N,N')}$ is a linear combination of nuclear response form factors $F_k^{(N,N')}$ and interference form factors $F_{k_1,k_2}^{(N,N')}$ that are calculated by sandwiching nuclear operators $k=M,\Sigma', \Sigma'', \Delta, \widetilde{\Phi}', \Phi''$ between nuclear states \cite{hax2}:
\begin{align}
	F_k^{(N,N')} (q^2) = \frac{4\pi}{2j_A +1} \sum_{J=0}^{2j_A+1} \langle j_A ||  k_J^{(N)} || j_A\rangle \langle  j_A ||  k_J^{(N')} || j_A \rangle \nonumber \\
	F_{k_1,k_2}^{(N,N')} (q^2) = \frac{4\pi}{2j_A +1} \sum_{J=0}^{2j_A+1} \langle j_A ||  k_{1J}^{(N)} || j_A\rangle \langle  j_A ||  k_{2J}^{(N')} || j_A \rangle \nonumber \\
\end{align}
The nuclear response form factors $F_k$ are not merely special cases of the operator form factors $F_{i,j}$.  Rather, they are well-defined quantities depending only on the physics of the target nucleus and can be approximated using standard nuclear physics techniques.  Numerical approximations to each of these form factors for common nuclei used in direct detection targets are calculated by Anand et. al using a standard shell model expanded over a set of Slater determinants and catalogued in \cite{hax1, hax3}.  For operators $\mathcal{O}_i$ and $\mathcal{O}_j$, the coefficients $a_{ijk}$ in the linear combination $F_{i,j}^{(N,N')} = \sum_{k} a_{ijk} F_k^{(N,N')}$ (where $k$ ranges over $\textrm{M}, \Sigma'', \Sigma', \Delta, \Phi'', \widetilde{\Phi}'$ and the two interference responses $(\textrm{M}, \Phi''), (\Sigma',\Delta)$) are simple products of WIMP and nucleon masses and spins.  These are catalogued in Appendix A.2 of \cite{hax1}.  All dependence on the dark matter physics (WIMP mass, WIMP spin, relative WIMP-target velocity, and so on) is built into the coefficients $a_{ijk}$.

\begin{figure}
	\vspace*{2mm}
	\includegraphics[width=1.0\columnwidth, trim={20mm 30mm 5mm 51mm}, clip]{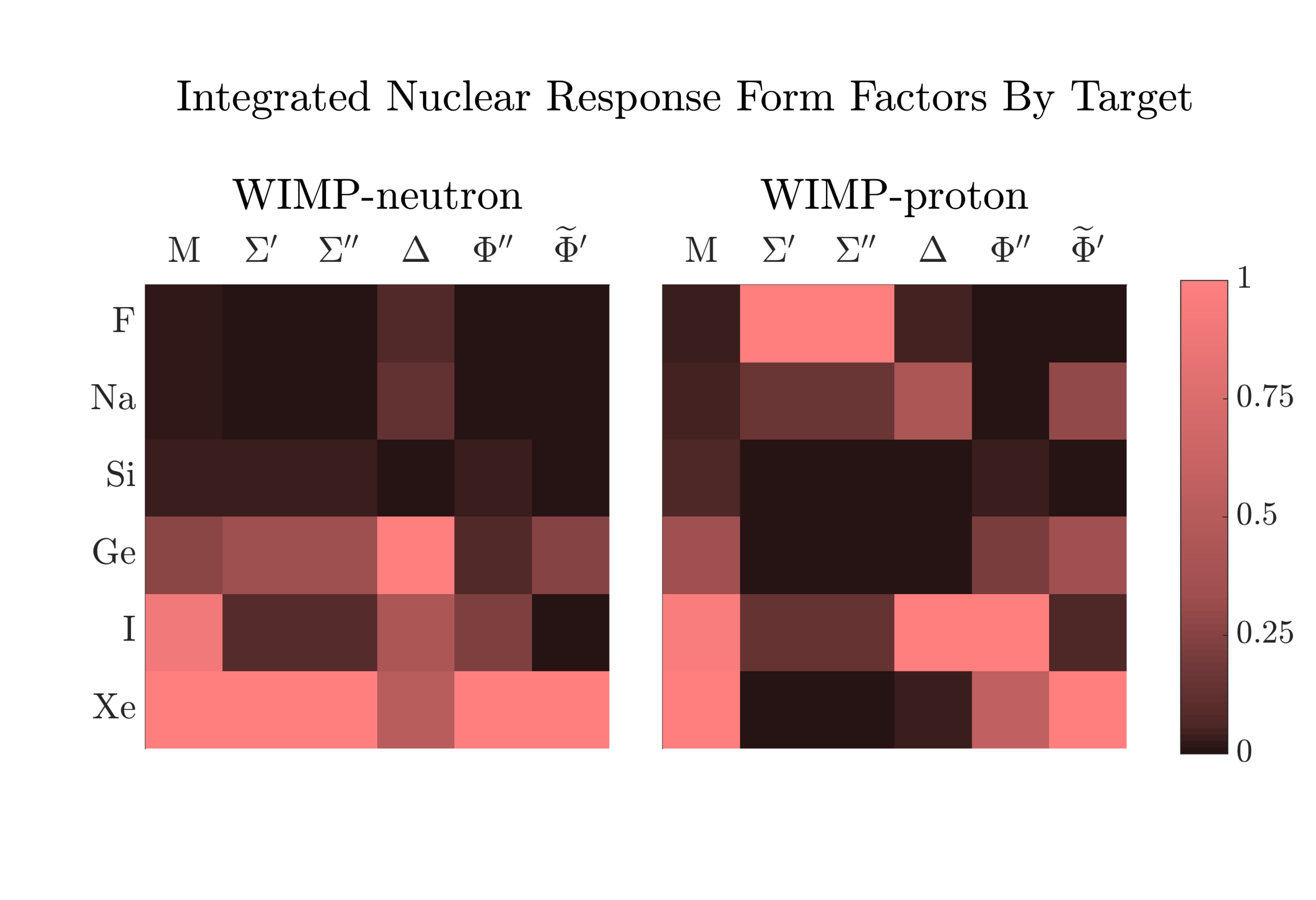}
	\vspace*{-6.5mm}
	\caption{The relative size of integrated nuclear form factors $\int_0^{100 \textrm{ MeV}}\frac{1}{2} q F_k(q^2)dq$ by target for $k= \textrm{M}, \Sigma'', \Sigma', \Delta,\Phi''$, and $\widetilde{\Phi}'$, adapted and expanded from Fig. 1 of \cite{hax1}.  The contribution of each isotope is weighted by natural abundance.  Each value is normalized by that of the element with the maximum integrated form factor. \label{fig:nuclearformfactorsbyelement}}
	\vspace*{-4mm}
\end{figure}

\begin{figure*}
	\centering
	\vspace*{8mm}
	\includegraphics[width=1.0\textwidth, trim={2mm 15mm 5mm 30mm}, clip]{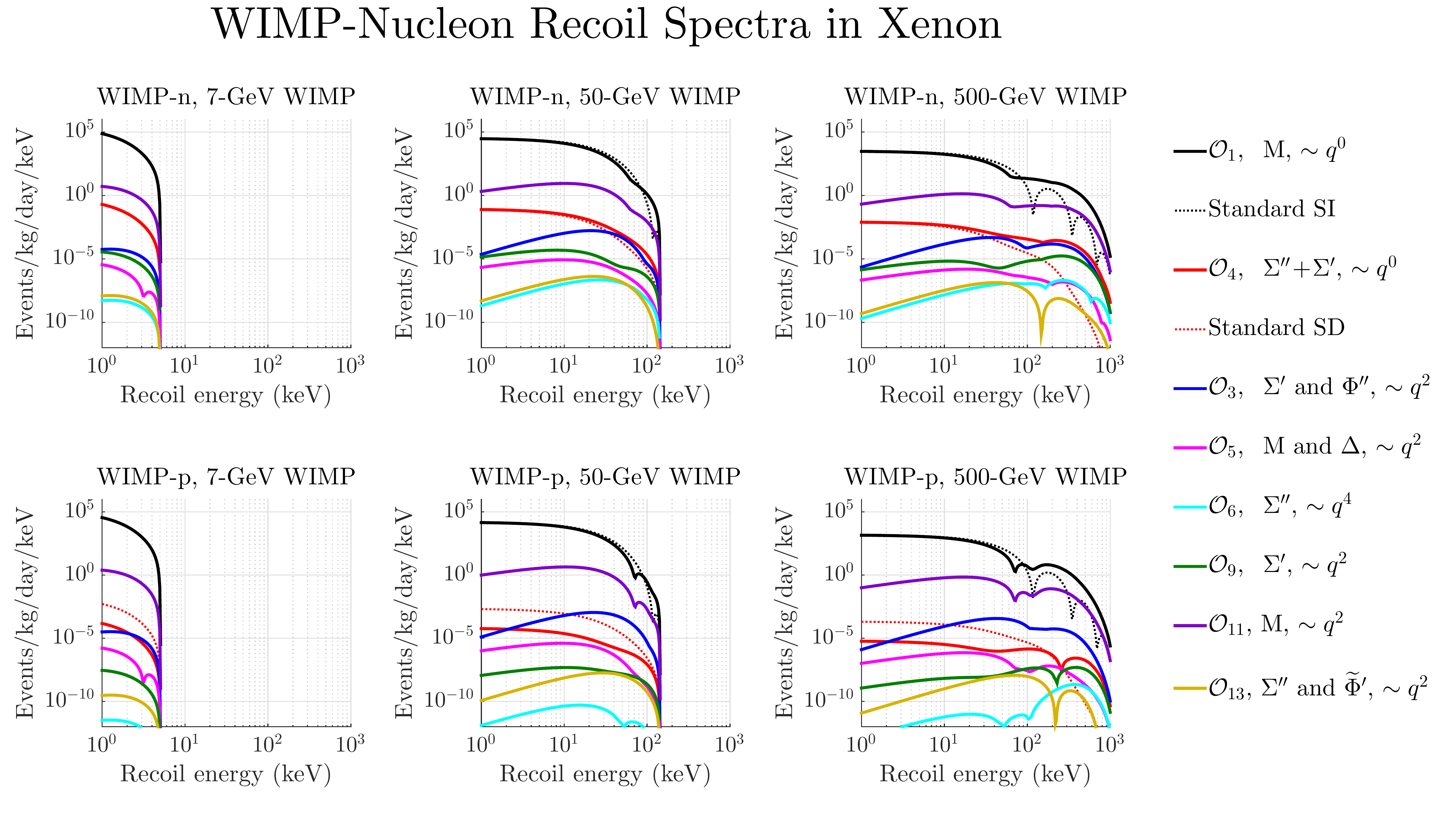}
	\vspace*{-4mm}
	\caption{WIMP-nucleon recoil spectra in xenon for a representative sample of EFT operators, weighted by the natural abundance of isotopes.  We assume a spin-$1/2$ WIMP that obeys a Maxwellian WIMP velocity distribution with a characteristic velocity $v_0=220$km/s and escape velocity $v_{\mathrm{esc}}=544$\,km/s.  The spectra are normalized so that the coupling constant $c_1^{(n,p)}$ produces a zero-momentum transfer WIMP-nucleon interaction cross section of $1\times10^{-36}$ cm$^2$ for operator $\mathcal{O}_1$. For the other operators $\mathcal{O}_i$, we set $c_i = c_1$.  For comparison, we also show the recoil spectra for the standard SI WIMP-nucleon interactions assuming a Helm form factor \cite{lewinsmith} and recoil spectra for the standard SD WIMP-nucleon interactions using the structure factors calculated in \cite{klos2013}. \label{fig:spectra}}
\end{figure*}

A comparison of form factors for common direct detection elements can be seen in Fig. \ref{fig:nuclearformfactorsbyelement}. Xenon has substantial sensitivity to the broadest set of interaction forms among practical WIMP targets; note that xenon is sensitive to WIMP-neutron interactions for all five nuclear responses, to SI and LSD/tLSD WIMP-proton interactions, and to a lesser extent to LD WIMP-proton interactions.  The scalability and low background levels of xenon TPCs like LUX can help offset their limited sensitivity to WIMP-proton SD interactions, so xenon is a good choice of target regardless of the nature of the interaction.    Even more sensitive coverage of WIMP-proton interactions can be achieved by the addition of fluorine and iodine WIMP targets. Complementarity between different target materials also helps ensure coverage of any ``blind spots" in xenon.  Such a scenario will occur if WIMPs are isospin-violating and interact predominantly with protons via operators that produce only $\Sigma''$, $\Sigma'$, and $\Delta$ responses.  

We consider the possible recoil spectra that can be produced in a xenon target.  The standard SI WIMP-nucleon recoil spectrum roughly follows a decaying exponential. For a 500-GeV WIMP undergoing an SI recoil in xenon, the vast majority of interactions (93\%) deposit $\leq$40\,keV; this is roughly the upper energy threshold used by LUX in previous WIMP searches.  Direct detection experiments have focused more on reducing the lower energy threshold than on studying higher-energy recoils.  However, the spectra for interactions governed by momentum-dependent operators are qualitatively flat or exhibit an upwards slope with respect to recoil energy out to tens or even hundreds of keV.   This motivates expanding the energy window, since momentum-dependent interactions are more likely to occur at high recoil energies, especially for higher-mass WIMPs. We show the expected NR spectrum from a section of operators and three example WIMP masses (low, medium, high) in Fig.\,\ref{fig:spectra}; here we make the standard assumptions that the WIMP velocities follow a Maxwell-Boltzmann distribution with characteristic velocity $v_0=220$\,km/s and escape velocity $v_{\mathrm{esc}}=544$\,km/s.  The features in the spectrum for each individual operator vary according to the nuclear responses that are produced in a given target material.  

\section{\label{section:LUX}The LUX Apparatus and First Dark Matter Search}

The LUX detector was a dual-phase xenon time projection chamber with an active mass of 250\,kg. The active xenon target was viewed by two arrays of 61 photomultiplier tubes (PMTs) each, one viewing the target from above and the other from below.  An incident WIMP interacting with the target xenon would appear as a nuclear recoil (NR) depositing energy up to $\mathcal{O}(100\,\mathrm{keV})$.  Such an energy deposition would produce both prompt scintillation light (S1) and ionization electrons.  The electrons were drifted upwards by an 
electric field established by a wire cathode grid, located above the bottom PMT array, and a gate grid, located just below the liquid surface of the xenon.  They were then extracted from the liquid surface and accelerated through the gas region by the electric field between the gate grid and an anode mesh underneath the top PMT array.  The rapidly-moving electrons luminesce in the gaseous xenon to produce a secondary (S2) scintillation signal, whose amplitude was proportional to the number of extracted electrons.  The S1 and S2 signals were both recorded by the PMTs.  The hit pattern of the S2 signal on the top PMT array yielded the $(x,y)$-position, while the difference in arrival time between the two pulses reflected the depth of the interaction.  This enabled the localization of the interaction site in 3D: the 1--$\sigma$ statistical resolution of reconstructed $(x,y)$ coordinates was 10\,mm at the S2 threshold, improving for larger S2 pulses proportional to S2$^{-1/2}$; the resolution of the reconstructed $z$ coordinate was better than 1\,mm\,\cite{luxrun3re2016}. Multiple-scatter events could be identified and rejected, and the active volume could be fiducialized to cut out external backgrounds which produced signals primarily in the outer regions of the active detector volume. The distribution of events in the S1-S2 plane could be used to reject nearly all background events in the form of gamma interactions and $\beta$ recoils. Further details of the LUX instrument including the design of the internals, the PMTs, the cryogenics and xenon circulation/purification system, and the trigger and readout electronics can be found in \cite{LUXinstrument2013}.

During mid-2013, LUX acquired 95 live-days of WIMP search data with a 145\,kg fiducial mass.  During this time, regular \textit{in situ} calibrations were performed with $^{83\mathrm{m}}$Kr, which was injected directly into the path of the xenon flow and allowed to disperse throughout the active volume.  As $^{83\mathrm{m}}$Kr is mono-energetic, these calibrations allowed the position-dependent detector response to be studied.  Additionally, a tritiated methane (CH$_3$T) source was developed for calibration of detector efficiencies and to define the broad-energy ER response\,\cite{LUXtritium2015}.  The detector's NR response was performed with a mono-energetic Adelphi deuterium-deuterium neutron generator, which formed a collimated neutron beam incident on the detector.  This calibration also facilitated the lowest-energy measurement of the LXe NR scintillation response\,\cite{LUXDD2015}, down to 1.1\,keV.  A minimum 90\% upper C.L.~limit of $6\times 10^{-46}$\,cm$^2$ to SI WIMP-nucleon interactions for WIMPs of mass 33\,GeV/$c^2$ was obtained with this data set\,\cite{luxrun3re2016}.  Additional analyses of these data set have been performed, placing limits on SD WIMP-nucleon interactions\,\cite{luxsd2016}, solar axions, dark-matter axion-like particles\,\cite{LUX_Run3_axionALP}, mirror dark matter\,\cite{Akerib:2019diq}, and sub-GeV dark matter\,\cite{Akerib:2018hck}.  Following the mid-2013 WIMP search campaign, a much longer data set of 332 live-days was acquired between September 2014 and May 2016.  These data, when combined with LUX's 2013 data, yielded a 90\% C.L.~upper limit on SI WIMP-nucleon interactions of $1.1\times 10^{-46}$\,cm$^2$ for 50\,GeV/$c^2$ WIMPs\,\cite{luxrun42016}, as well as updated limits SD WIMP-nucleon couplings\,\cite{LUX_SD_Run3_Run4}. The 2014-2016 LUX operation and analysis was more complex than in 2013 because of electric charging of the PTFE walls, which substantially distorted the drift field. In the next section, we use predicted EFT nuclear recoil spectra to calculate limits on operator coupling constants from the LUX 2013 dataset, which is simpler to analyze due to the roughly constant drift field. 


\section{LUX Limits on WIMP-Nucleon Effective Field Theory Interactions}
In this study, the detector parameters and background models used are consistent with those used in the analysis of LUX's first results\,\cite{luxrun3re2016}.  LUX's yields are quantified by the gain factors $g_1 = \langle \textrm{S1}_\textrm{corrected}\rangle/n_{\gamma} = 0.117\pm0.003$\,phd/photon and $g_2 = \langle \textrm{S2}_\textrm{corrected} \rangle  /n_e=  12.1 \pm 0.8$\,phd/electron, where $n_{\gamma}$ is the absolute number of photons produced in an interaction, $n_{e}$ is the absolute number of electrons, and $\langle \textrm{S1}_\textrm{corrected} \rangle$ and $\langle \textrm{S2}_\textrm{corrected} \rangle$ are the average of the measured S1 and S2 signals after correcting for position-dependent variations\,\cite{luxrun3re2016}.  Pulse areas are measured in units of detected VUV photons (phd) rather than the more traditional units of photoelectrons (phe), in order to incorporate the contribution of double-photoelectron emission from the PMT photocathodes\,\cite{faham2015}.  In effect, $g_1$ measures the average probability to detect an S1 scintillation photon; $g_2$ measures the average number of detected S2 photons per electron leaving an interaction vertex.

\begin{figure}
	\vspace*{0mm}
	\includegraphics[width=1.0\columnwidth]{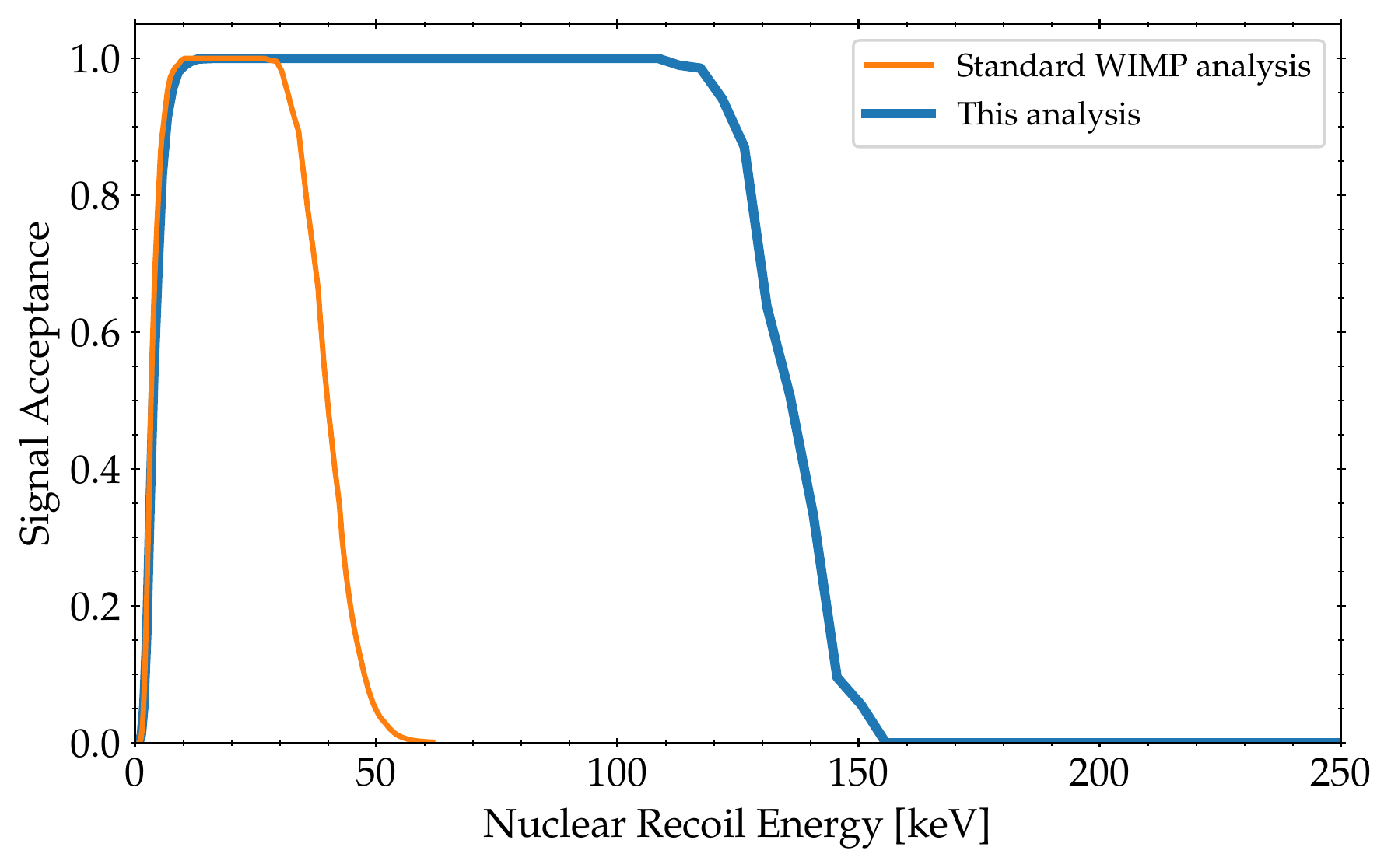}
  	\caption{LUX detector efficiencies used to define the search window for WIMPs undergoing EFT interactions with xenon nuclei in the LUX target.  The extension of the search window to higher energies is important as many EFT operators have significant signal response at energies much higher than in the standard WIMP search.  The background-discrimination cut ($\sim$50\% acceptance) is not included in this plot. \label{fig:det_eff}}
\end{figure}

\subsection{Extending the LUX WIMP Search Window}

The search window chosen for the standard LUX WIMP analyses gives sensitivity to xenon nuclear recoils (NRs) up to roughly 40\,keV.  However, the recoil spectra from many effective operators, seen in Fig.\,\ref{fig:spectra}, have significant amplitude at energies above this range. Unfortunately, studies in the literature on LXe's response to NRs above $\mathcal{O}$(100\,keV) are limited.  An \textit{in situ} measurement of LUX's response to a monoenergetic neutron source gives an endpoint NR energy of 74\,keV\,\cite{LUXDD2015}.  Measurements exist in the literature of LXe's NR response extending to much higher energies\,\cite{Sorensen:2011bd}, the highest coming from the endpoint of AmBe neutrons, at 333\,keV.  The Noble Element Simulation Technique (NEST) software package\,\cite{NESTweb} utilizes the world's literature on LXe's NR response (including these measurements at high energy) and can interpolate across the range of energies of interest in this study.

An additional challenge encountered when extending the WIMP-search window to higher energies is the presence of residual $^{83\textrm{m}}$Kr decays.  This isotope was periodically injected into the LXe for calibration purposes, and trace amounts often remained in the fiducial volume after the conclusion of such a calibration.  The source deposits 41.5\,keV mainly in the form of internal-conversion and Auger electrons; these electronic recoils (ERs) produce scintillation and ionization signals that are similar to that of a 210\,keV NR (LXe responds to ERs and NRs differently).  As a result, residual $^{83\textrm{m}}$Kr presents no background for a standard WIMP search, but it can potentially impact studies which search for higher-energy NRs.  To avoid this, we restrict our data sample to events whose energy is less than 5$\sigma$ below the peak observed from $^{83\textrm{m}}$Kr, which corresponds roughly to 150\,keV NRs.  The low-energy NR response of LUX has been measured down to 1.1\,keV, below which we conservatively assume LXe has no response.  The signal acceptance as a function of NR energy is shown in Fig.\,\ref{fig:det_eff}, as well as a comparison to that from the standard WIMP search.

\subsection{WIMP-Neutron and WIMP-Proton Limits} \label{section:swoosh}

\begin{figure}[t!]
  	\centering
    \includegraphics[width=1.0\columnwidth]{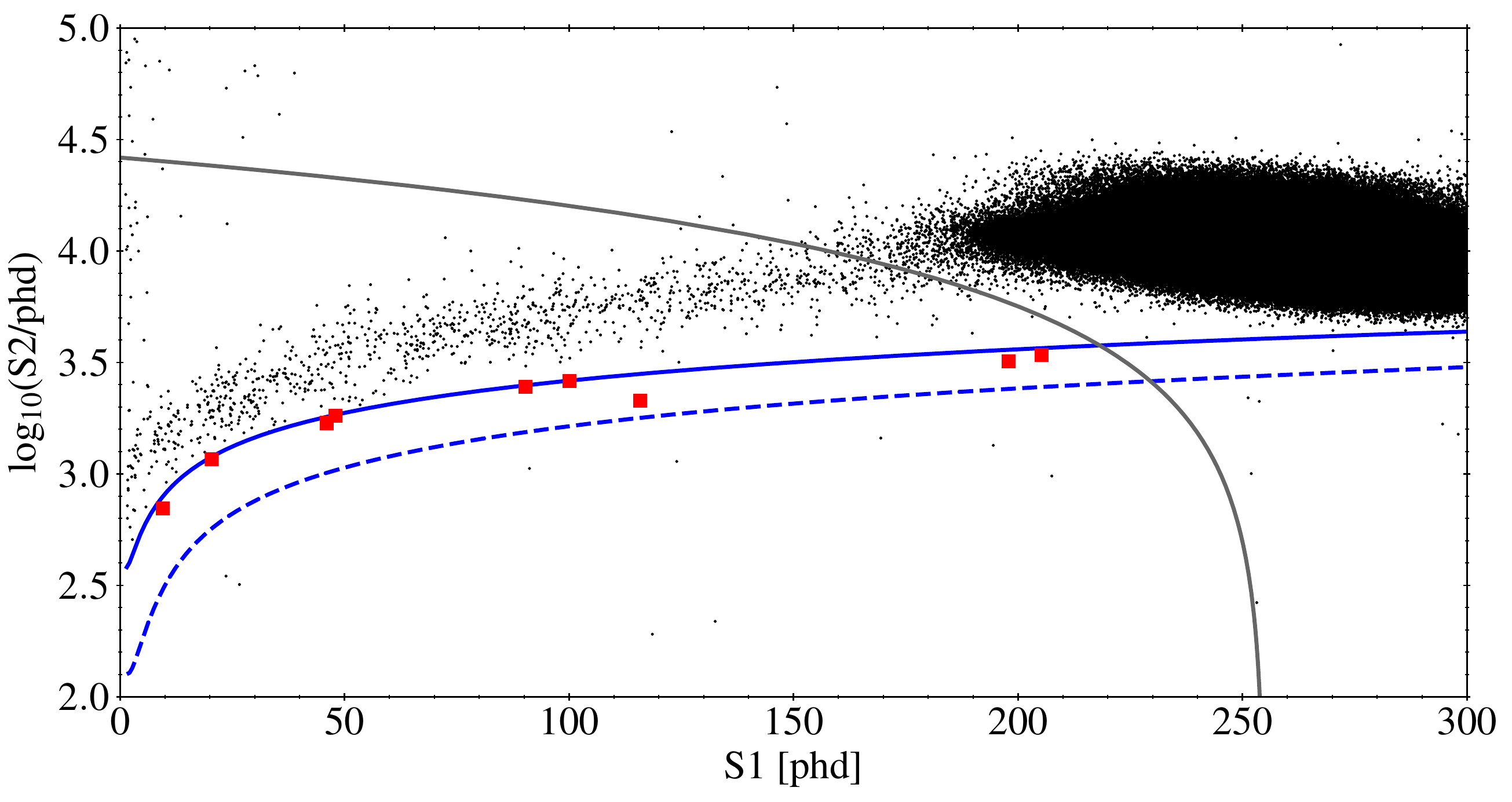}
  	\caption{All events in the 95-liveday LUX 2013 WIMP search dataset in log$_{10}$(S2) vs. S1 parameter space, for a 118-kg fiducial volume, after all cuts.  The solid blue curve indicates the median of the NR band as predicted by NEST, which has been tuned to NR calibration data (this differs slightly from LUX standard WIMP analysis of the same data, which used the NR band as measured with calibration data).  The dashed blue line shows the corresponding median$-3\sigma$ contour.  Events that lie between the solid and dashed blue lines are highlighted as red squares.  The gray curve is a contour of constant energy (corresponding to NRs of 150 keV), and defines the upper boundary of the search window; it is chosen to remove events originating from the decay of $^{83\mathrm{m}}$Kr (dense population of black points).  Overall, nine events appear in the signal region.\label{fig:WSgolden}}
\end{figure}

LUX's first WIMP-search run consisted of 95 livedays of data after removing periods of detector instability.  A conservative fiducial radius of $r<18$\,cm is chosen in the present work in order to avoid background events originating from the decay of radon daughters implanted on the detector wall.  This fiducial radius, together with a $z$-cut of 8.45 cm $<$ $z$ $<$ 48.62 cm (as measured from the bottom PMT windows) corresponds to a fiducial mass of 118\,kg. This fiducial volume is the same as used for previous LUX searches for axions and sub-GeV dark matter using the 2013 data set. Basic data quality cuts are applied to eliminate pathologies such as long tails of pulses following large S2s.  In order to remove the majority of events arising from electronic recoils, we accept only events whose S2/S1 ratio lies below the median expected for a NR source.  Though the exact value of this median S2/S1 curve depends on the energy spectrum of NR, we define the signal-acceptance region here based on the median S2/S1 value of a broad-energy source.  Exact signal acceptance values, which will differ from 50\%, are calculated individually for each WIMP mass and operator. Additionally, we cut events with an S2 lying greater than $3\sigma$ below the median of the NR band.  The events passing all of these cuts are shown in Fig.\,\ref{fig:WSgolden}.  After all cuts, there were a total of nine events observed in the WIMP search region.  

\begin{figure}[t]
  	\centering
    \includegraphics[width=1.0\columnwidth]{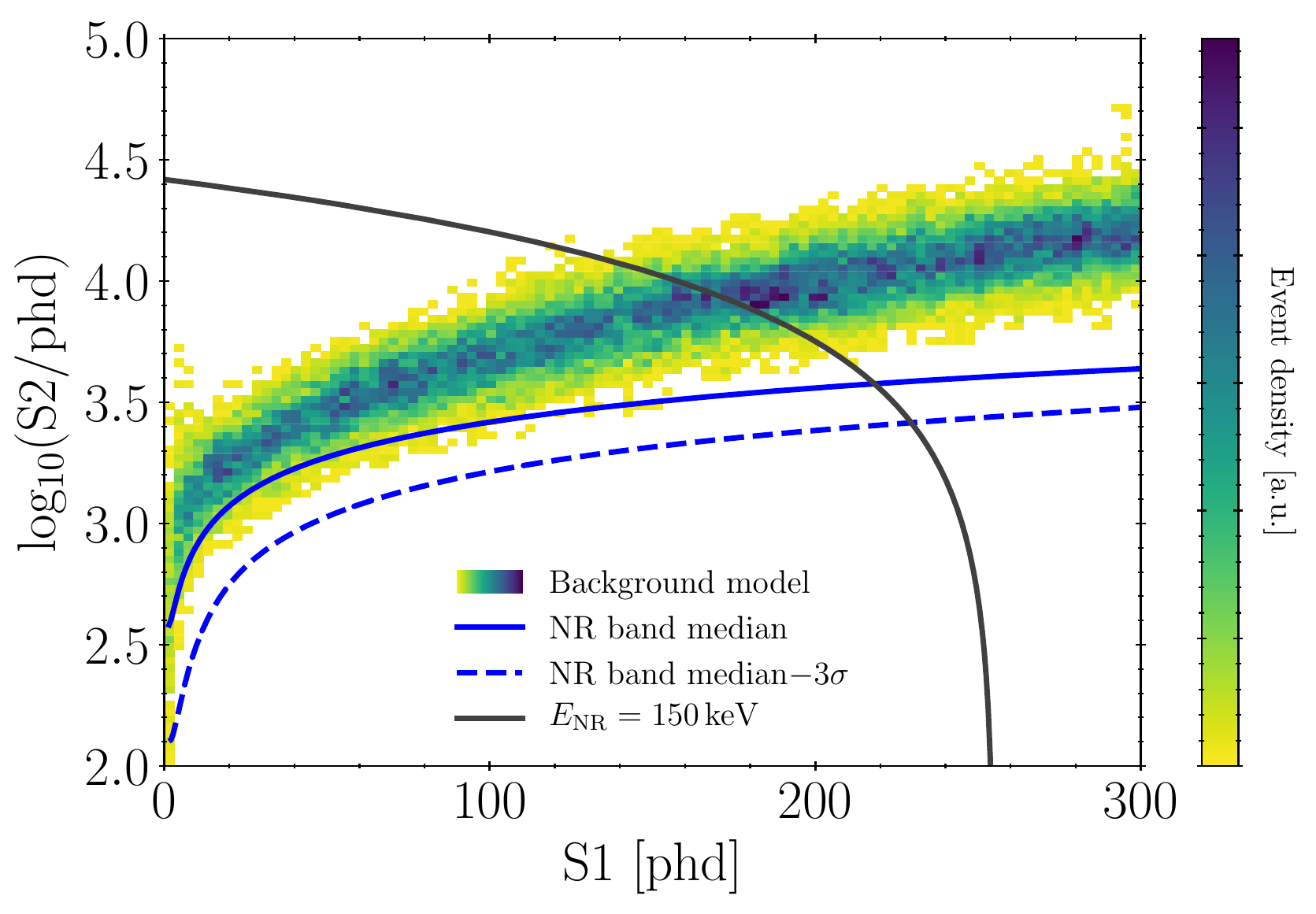}
    \caption{\label{fig:bg}Simulated ER background event density in LUX for a 118-kg fiducial volume (18-cm radius).  The median (solid blue) and median$-3\sigma$ (dashed blue) contours of the NR band are shown.  The WIMP signal region is defined as the region between these two blue contours, up to the gray contour which represents a nuclear recoil energy (`$E_{\mathrm{NR}}$') of 150\,keV.}
\end{figure}

The background model used in previous LUX WIMP searches, derived from the Geant4-based LUXSim package\,\cite{LUXSim2012}, comprises ER signals from $^{232}$U/$^{238}$Th/$^{40}$K decays in solid detector materials, as well as dissolved $^{85}$Kr, $^{37}$Ar, and $^{214}$Pb (a $^{222}$Rn daughter) decays in the LXe bulk. When combined, these many sources conspire to produce a background ER spectrum roughly flat in energy.  The simulated spectrum in the region of interest in the present work is shown in Fig.\,\ref{fig:bg}. The events below the $-3\sigma$ contour of the NR band seen in the data (Fig.\,\ref{fig:WSgolden}) contribute negligibly to the standard, low-energy WIMP search, yet constitute a distinct departure from these expected backgrounds at higher energies.  It is expected that these events arise from misidentified multiple gamma-ray scatters; this can happen when, for example, the gamma ray scatters once in the active region, and once in a charge-insensitive region (e.g.~below the cathode grid). In such a scenario, only a single S2 pulse is observed (leading to the misidentification), while the S1 pulses from both scatters are reconstructed together in time.  A precise model for this class of backgrounds is challenging and is currently under investigation.  Because of this, we choose to forgo background subtraction, and instead derive an upper limit treating all events as possible signal.

\begin{figure}[h!]
	\includegraphics[width=1.0\columnwidth]{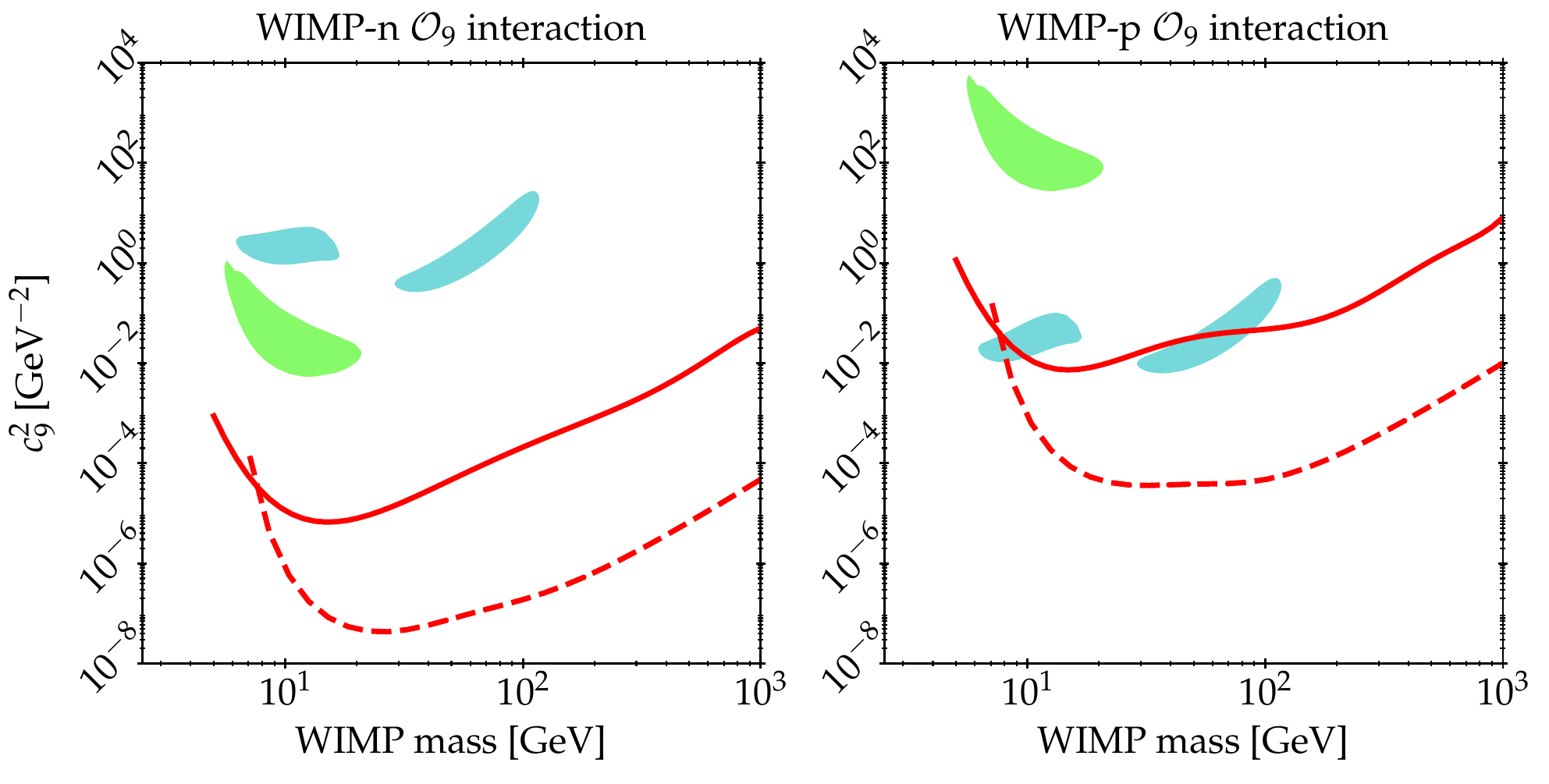}
	\vspace*{1mm}
  	\makebox[\columnwidth][l]{\includegraphics[width=1.0\columnwidth]{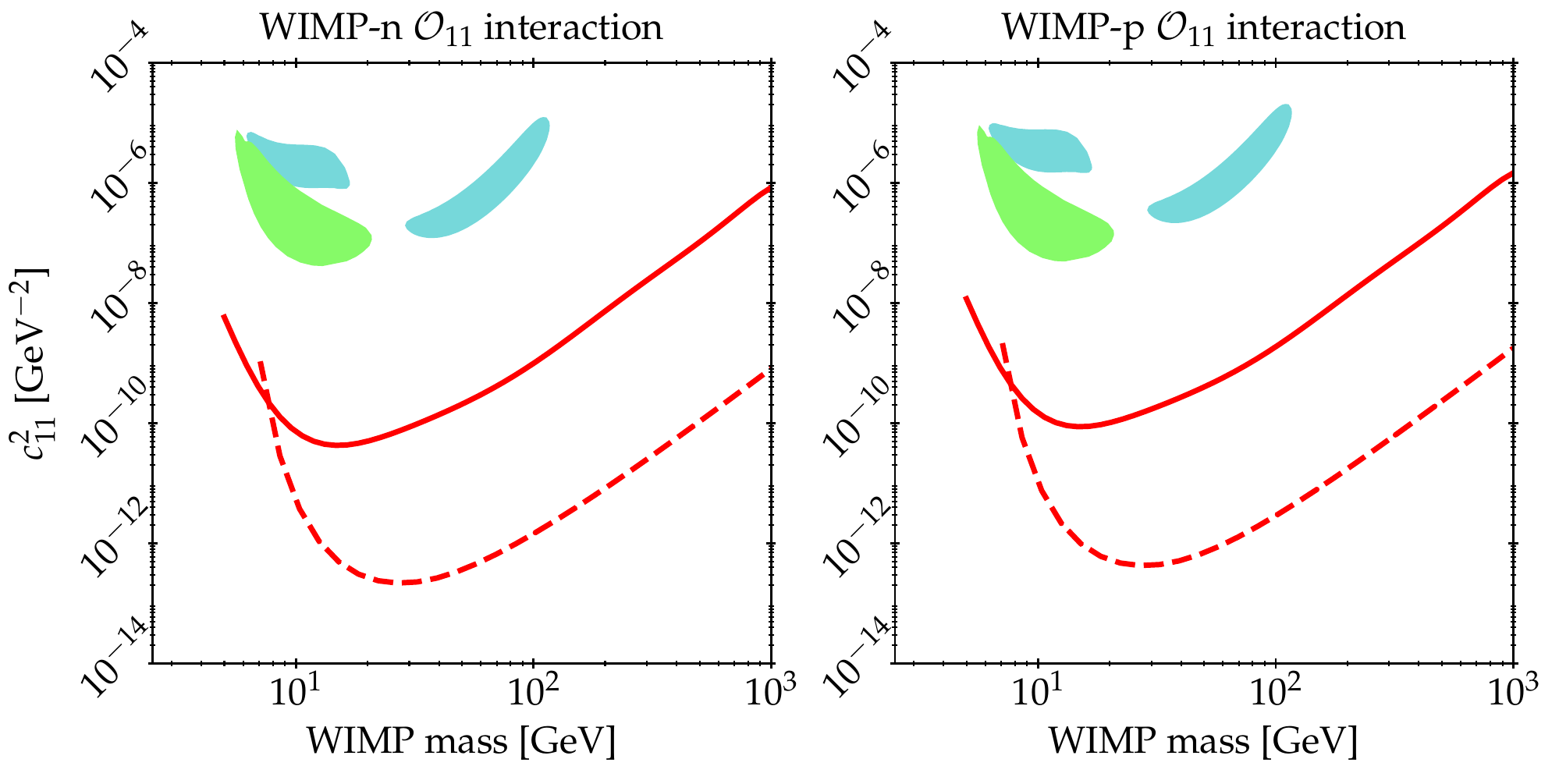}}
	\vspace*{1mm}
  	\makebox[\columnwidth][l]{\includegraphics[width=1.0\columnwidth]{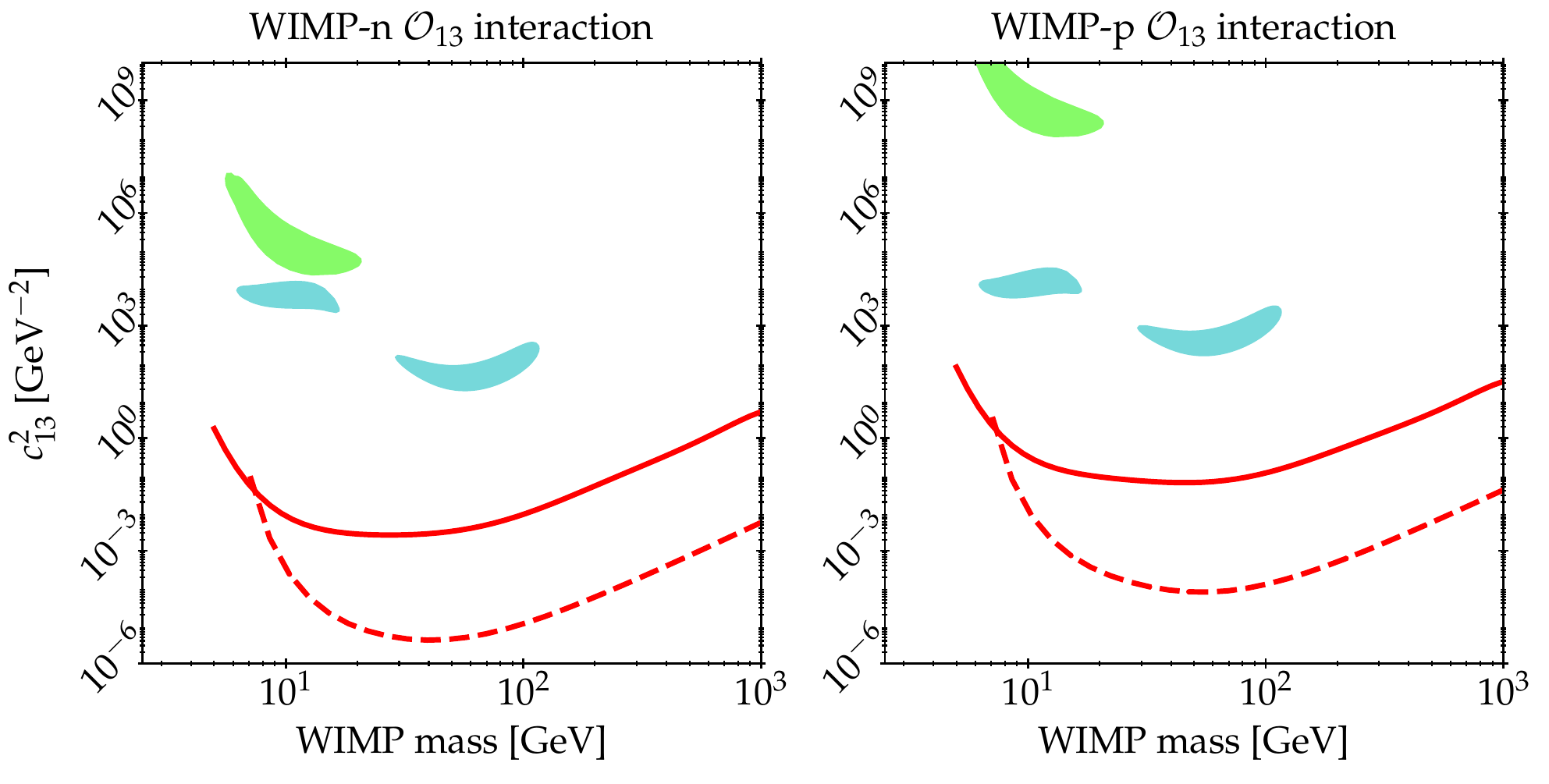}}
	\vspace*{-5mm}
  	\caption{Limits on the WIMP-neutron and WIMP-proton coupling constants associated with an example selection of EFT operators $\mathcal{O}_1$ (the standard SI operator, no dependence on momentum $q$), $\mathcal{O}_3$ (generates $\Sigma'$ and $\Phi''$ responses proportional to $q^2$), $\mathcal{O}_4$ (the standard SD operator, no dependence on $q$), $\mathcal{O}_5$ (generates M and $\Delta$ responses proportional to $q^2$), $\mathcal{O}_6$ (generates a $\Sigma''$ response proportional to $q^4$), $\mathcal{O}_9$ (generates a $\Sigma'$ response proportional to $q^2$), $\mathcal{O}_{11}$ (generates an M response proportional to $q^2$), and $\mathcal{O}_{13}$ (generates $\Sigma''$ and $\widetilde{\Phi}'$ responses proportional to $q^2$). \label{fig:swoosh}}
\end{figure}

Yellin describes in detail three techniques for setting one-sided exclusion limits in the presence of un-modeled backgrounds\,\cite{Yellin:2002xd}.  These techniques perform an un-binned comparison between observed events and a given signal model, considering both signal amplitude and distribution in one variable.  In scenarios where the observed data show marked departure from an expected signal, derived exclusion limits are in general stronger than would be obtained via a simple counting analysis.  We utilize Yellin's $p_{\mathrm{max}}$ test statistic, setting limits on individual effective coupling constants, separately for neutron and proton couplings.  For example, the upper limit on the coupling constant $c^{(n)}_9$, associated with pure-neutron coupling via $\mathcal{O}_9$, is found by assuming $c^{(p)}_9 = c^{(p,n)}_{i\neq9} = 0$.  A selection of these limits are shown in Fig.\,\ref{fig:swoosh}.  The signal models, shown in Fig.\,\ref{fig:spectra}, are compiled by averaging over xenon isotopes, weighted by natural abundance. Numerical form factors for each of the operators $\mathcal{O}_{1-11}$ are obtained from\,\cite{hax1}, while form factors for the exotic operators $\mathcal{O}_{12-15}$ are calculated using the Mathematica package DMFormFacor created by N.~Anand \textit{et al.}\,\cite{dmformfactor, hax3}. We assume a spin-$\frac{1}{2}$ WIMP, although the formulae in Section\,\ref{section:EFTframework} accommodate a WIMP of any spin. Operators which involve the spin of the nucleus ($\vec{S}_N$) may couple only to $^{129}$Xe and $^{131}$Xe, whose spins are dominated by an unpaired neutron; as a result, the pure-proton limits for these operators are relatively weak.

\begin{figure}[h!]
	\centering
	\vspace*{0mm}
    \includegraphics[width=1.0\columnwidth]{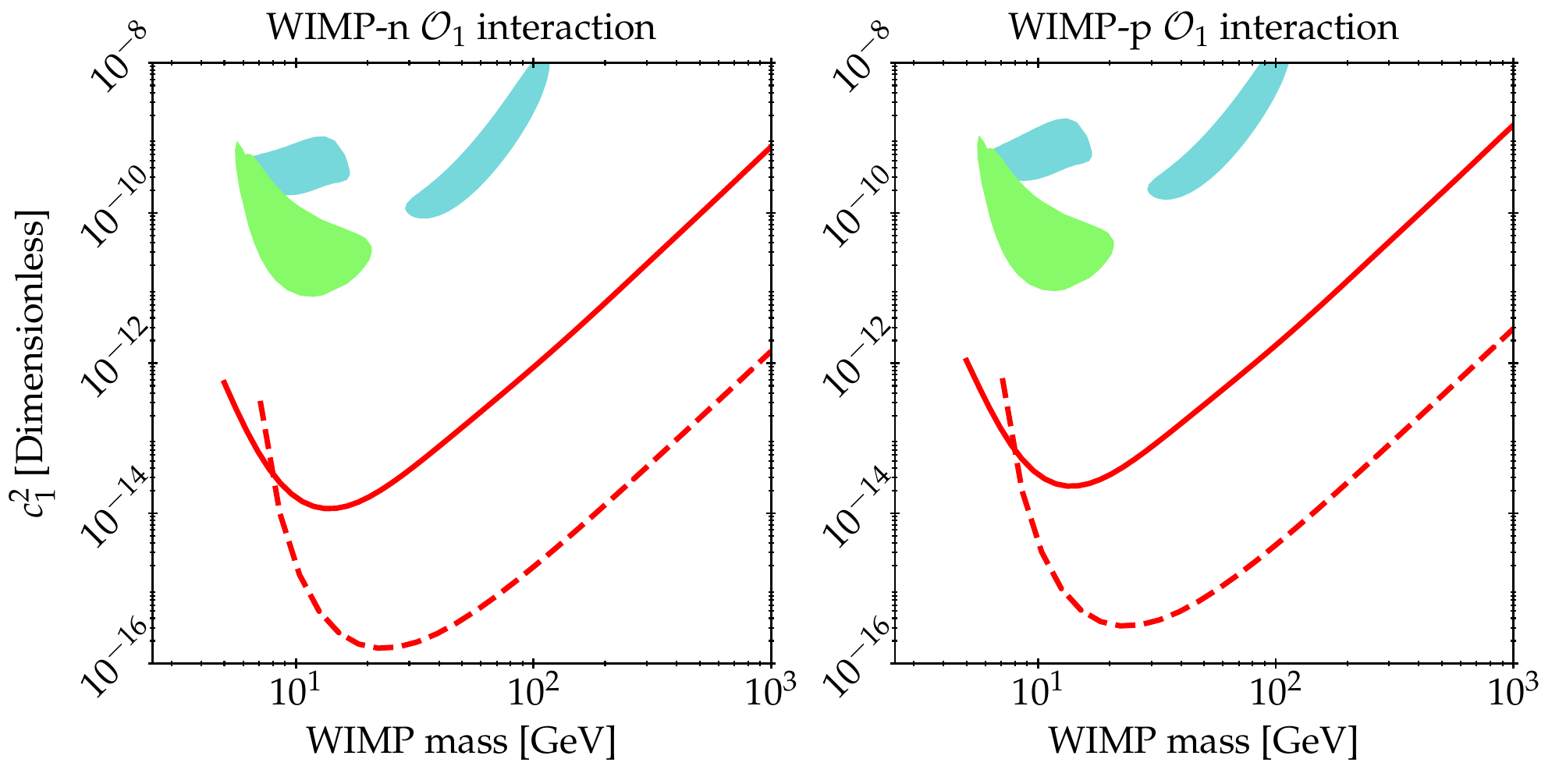}
	\vspace*{1mm}
  	\includegraphics[width=1.0\columnwidth]{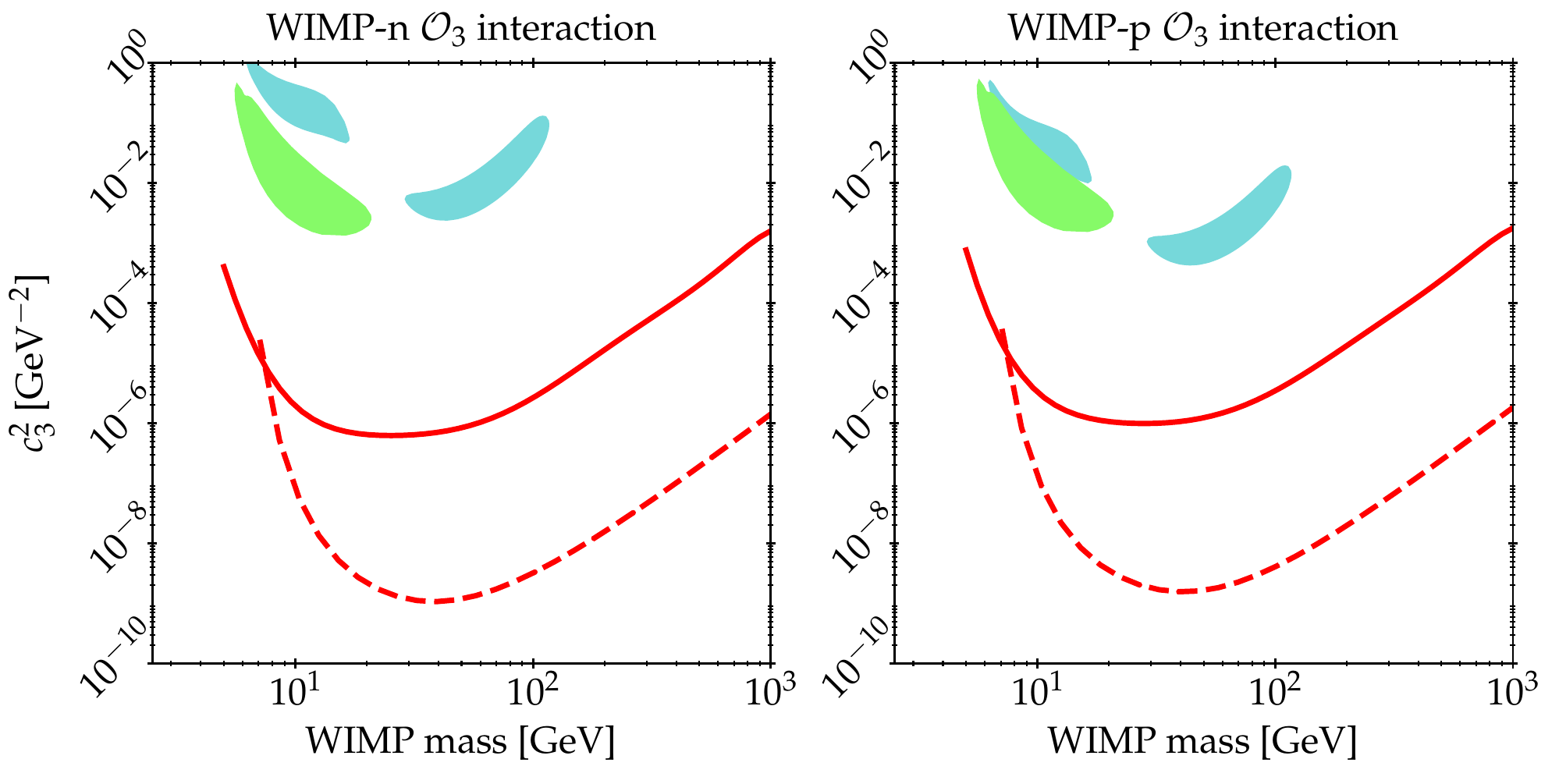}
	\vspace*{1mm}
  	\includegraphics[width=1.0\columnwidth]{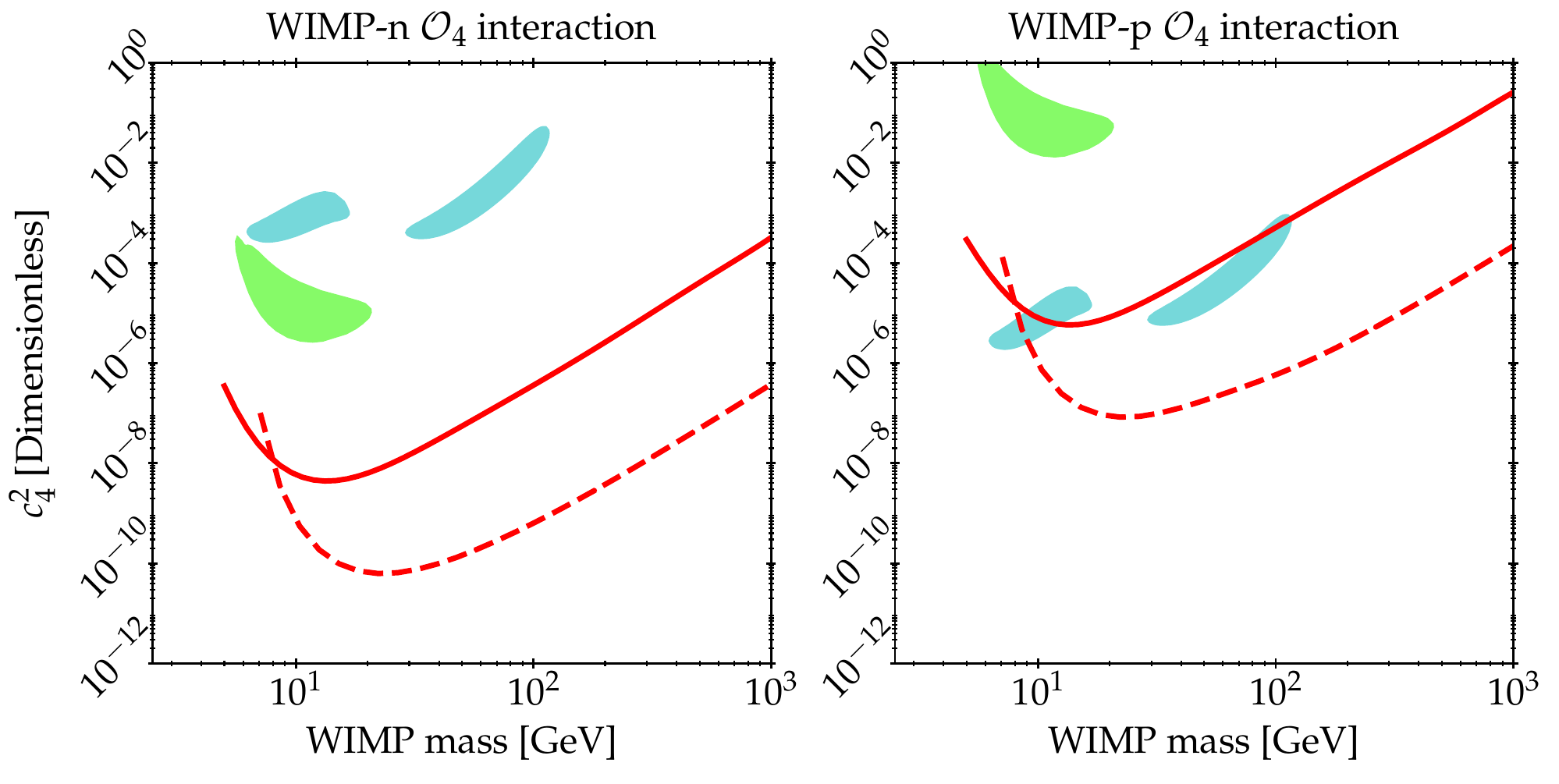}
	\vspace*{1mm}
  	\includegraphics[width=1.0\columnwidth]{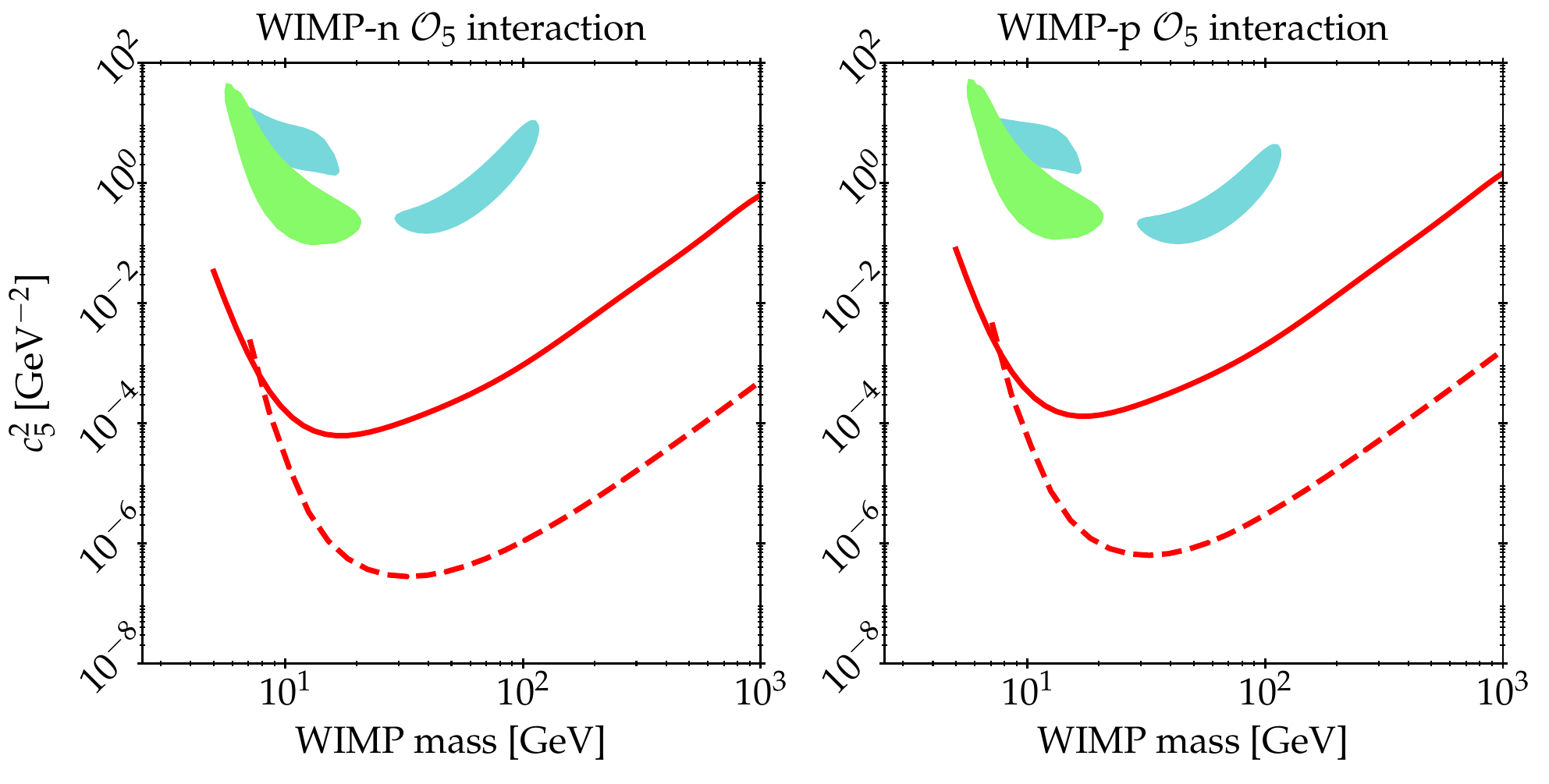}
	\vspace*{1mm}
 	\includegraphics[width=1.0\columnwidth]{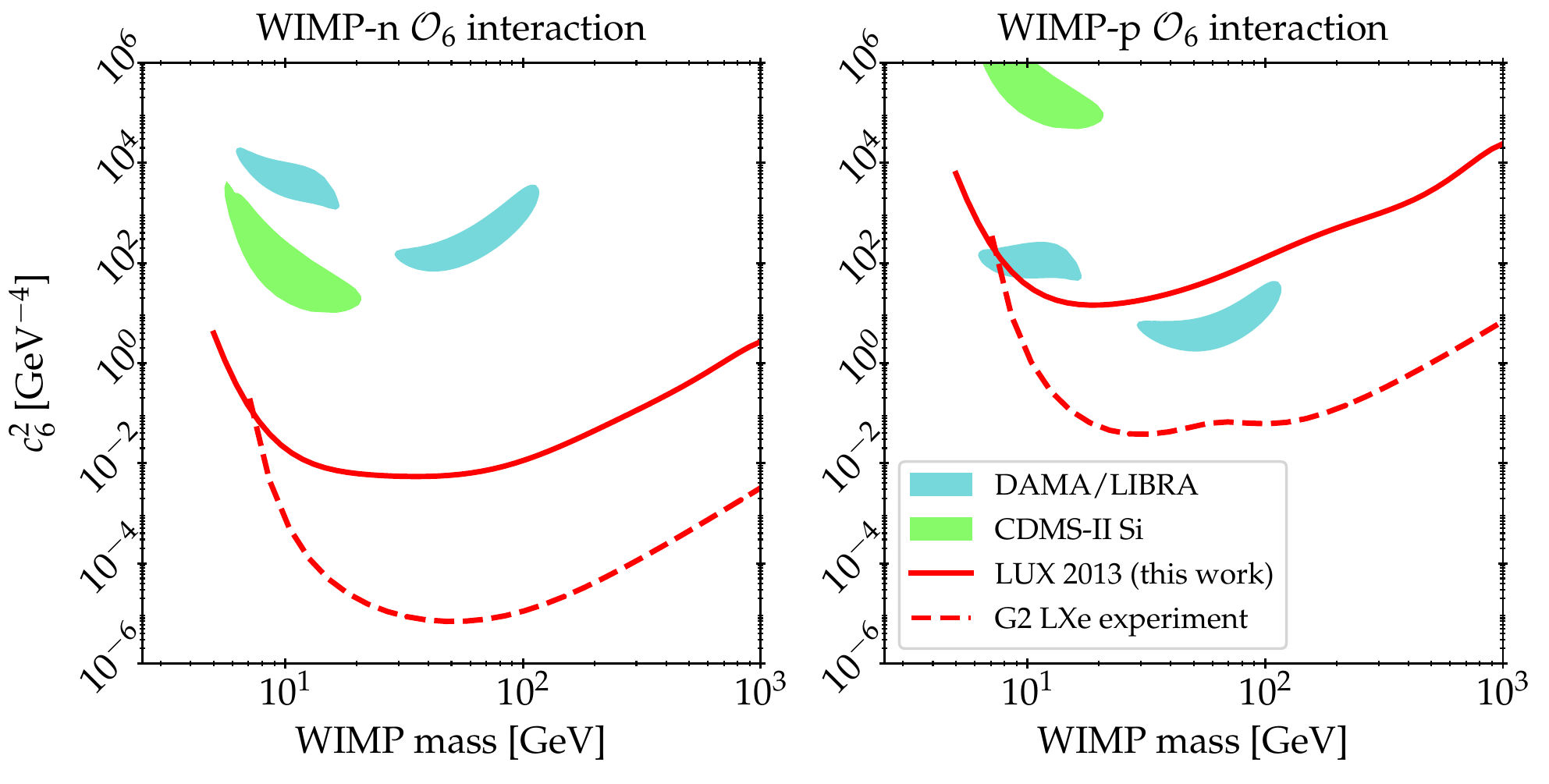}
	\vspace*{-12mm}
\end{figure}

We compute favored regions of parameter space for two experiments reporting an excess of WIMP-like events in their data.  The DAMA/NaI and DAMA/LIBRA experiments, which consist of a combined 2.46 ton-years of data taken with an array of sodium iodide crystals, report a 12.9$\sigma$ combined annual modulation result consistent with a WIMP\,\cite{Bernabei:2018yyw}.  The CDMS-II experiment consisted of an array of germanium and silicon detectors that operated from 2003 to 2008.  No evidence of WIMP-like recoils were observed in the germanium detectors; however, an exposure of 140.2 kg-days of the silicon detectors (23.4 kg-d after all selection cuts) yielded a total of 3 observed nuclear recoil events (0.54 background events expected) in the energy range 7-100\,\cite{cdmssi2013}.  The CDMS-II Si result favored a WIMP explanation over a background-only hypothesis, but was not interpreted by the collaboration as a WIMP discovery.  Nevertheless, we include the CDMS-II Si-favored region in our analysis.  For both DAMA/LIBRA and CDMS-II Si, we convert the reported 90\% confidence regions in the $\sigma_{\textrm{SI}}$-WIMP mass plane to a region in $c_{i}^{(n,p)2}$ vs. WIMP mass space using the ratio of the integrated SI recoil spectrum to the integrated $\mathcal{O}_i$ recoil spectrum over the relevant energy window, taking into account the reported detector efficiencies in Fig.\,26 of \cite{damaapparatus2008} for DAMA/LIBRA and in Fig.\,1 of \cite{cdmssi2013} for CDMS-II Si.

Finally, we estimate EFT sensitivities of a future ``Generation-2'' multi-tonne LXe experiment. We take the parameters of the upcoming LUX-ZEPLIN (LZ) experiment\,\cite{Akerib:2019fml} as a baseline for this estimation: 5600\,kg fiducial mass, 1000\,days livetime.  The LZ experiment predicts a background of 6.18\,events in an NR range of 6--30\,keV\,\cite{Akerib:2018dfk}, after a 50\% signal-acceptance cut to veto ERs.  We wish to estimate how such a background prediction would scale to an extended energy window (as is done in this work) up to 150\,keV. The ER and NR bands in LXe diverge at high energies, with negligible overlap for NRs above $\sim$75\,keV; because of this, we assume this 6.18 predicted background events scales linearly when extending the signal window up to 75\,keV, with no further background contribution for NRs in the range 75--150\,keV.  We estimate 90\%-C.L. sensitivity of such a G2 experiment with this predicted number of background counts using the frequentist methods of Feldman and Cousins\,\cite{feldmancousins1998}.  These predicted G2 sensitivities are shown as red-dashed curves in Fig.\,\ref{fig:swoosh}.

The XENON100\,\cite{Aprile:2017aas} and PandaX-II\,\cite{Xia:2018qgs} collaborations have released results on EFT studies, also using LXe as a detector target.  However, these collaborations have chosen to perform their analyses in ways that make direct comparison to our results (and to each other) infeasible: XENON100 presents results based on pure-isoscalar and pure-isovector couplings (rather than pure-proton/pure-neutron as we have done here); PandaX-II also chooses to set constraints on isoscalar/isovector couplings, but in addition uses a relativistic EFT framework (rather than the nonrelativisitic framework used by XENON100 and us).  Because of this, we show no comparisons to XENON100 and PandaX-II in this work, and note the need for the community to come to a consensus in future EFT studies which will facilitate such comparisons. Despite these different approaches, the results from all three experiments are likely to offer similar sensitivity.

\section{Conclusion}

Typical dark matter direct detection analyses, by assuming strictly nonrelativistic momenta, greatly restrict the space of possible WIMP-nucleon interactions that can be investigated.  Motivated by the unresolved tension between experimental results and the absence of a definitive positive detection despite great advances in detector sensitivities, many dark matter searches are beginning to broaden the scope of their approaches.  This includes searching for extremely light or heavy WIMPs, isospin-violating WIMPs, or WIMPs that undergo exotic interactions with nuclei.  The effective field theory summarized here enables the exploration of a rich parameter space in a more model-independent way, allowing for WIMP-nucleon interactions of greater complexity.  Underneath this framework, the possible signal models produce very different spectral shapes than the standard SI or SD interactions, motivating the expansion of search windows to encompass higher-energy nuclear recoil events.  In addition to SI and SD responses, novel nuclear responses such as angular-momentum-dependent (LD) and scalar or tensor spin-orbit (LSD) responses can be produced.  The magnitude of each of these nuclear responses varies between targets, so it is possible that event rates are greatly suppressed in one experiment compared to another and that an array of complementary targets are needed to rule out all possible WIMP scenarios.

Here, we have significantly widened the LUX SI WIMP search window to higher energies, and we have used the resulting data set to calculate limits on the interaction strengths of all 14 EFT operators.  For all scenarios except those where WIMPs interact with nuclei primarily through a WIMP-proton operator that produces only spin-dependent responses, LUX produces the tightest WIMP-nucleon interaction constraints.  Future generations of direct detection experiments should therefore consider both odd-neutron and odd-proton targets to ensure full coverage of all available parameter space.

An effort is now underway to streamline and incorporate effective field theory signal models into the profile likelihood ratio analysis detailed in \cite{luxrun42016} to search for exotic dark matter interactions in the full LUX exposure.  We anticipate that the additional sensitivity of next generation detectors, coupled with the more comprehensive analytical framework discussed in this paper, will help clarify the nature of dark matter to a much greater degree than possible in previous WIMP analyses.

\section{Acknowledgements}
The research supporting this work took place in whole or in part at the Sanford Underground Research Facility (SURF) in Lead, South Dakota. Funding for this work is supported by the U.S. Department of Energy, Office of Science, Office of High Energy Physics under Contract Numbers DE-SC0019066, DE-SC0020216, DE-FG02-08ER41549, DE-FG02-91ER40688, DE-FG02-95ER40917, DE-FG02-91ER40674, DE-NA0000979, DE-FG02-11ER41738, DE-SC0015535, DE-SC0006605, DE-AC02-05CH11231, DE-AC52-07NA27344, and DE-FG01-91ER40618. This research was also supported by the U.S. National Science Foundation under award numbers PHYS-0750671, PHY-0801536, PHY-1004661, PHY-1102470, PHY-1003660, PHY-1312561, PHY-1347449; the Research Corporation grant RA0350; the Center for Ultra-low Background Experiments in the Dakotas (CUBED); the South Dakota School of Mines and Technology (SDSMT). LIP-Coimbra acknowledges funding from Funda\c{c}\~{a}o para a Ci\^{e}ncia e a Tecnologia (FCT) through the project-grant PTDC/FIS-PAR/28567/2017. Imperial College and Brown University thank the UK Royal Society for travel funds under the International Exchange Scheme (IE120804). The UK groups acknowledge institutional support from Imperial College London, University College London and Edinburgh University, and from the Science \& Technology Facilities Council for PhD studentship ST/K502042/1 (AB). The University of Edinburgh is a charitable body, registered in Scotland, with registration number SC005336. The assistance of SURF and its personnel in providing physical access and general logistical and technical support is acknowledged.

\bibliographystyle{apsrev4-1}
\bibliography{eftpaper}

\end{document}

%% file: lux-tex-author-list_include.tex
\author{D.S.~Akerib} \affiliation{SLAC National Accelerator Laboratory, 2575 Sand Hill Road, Menlo Park, CA 94205, USA} \affiliation{Kavli Institute for Particle Astrophysics and Cosmology, Stanford University, 452 Lomita Mall, Stanford, CA 94309, USA} 
\author{S.~Alsum} \affiliation{University of Wisconsin-Madison, Department of Physics, 1150 University Ave., Madison, WI 53706, USA}  
\author{H.M.~Ara\'{u}jo} \affiliation{Imperial College London, High Energy Physics, Blackett Laboratory, London SW7 2BZ, United Kingdom}  
\author{X.~Bai} \affiliation{South Dakota School of Mines and Technology, 501 East St Joseph St., Rapid City, SD 57701, USA}  
\author{J.~Balajthy} \affiliation{University of California Davis, Department of Physics, One Shields Ave., Davis, CA 95616, USA}  
\author{A.~Baxter} \affiliation{University of Liverpool, Department of Physics, Liverpool L69 7ZE, UK}  
\author{E.P.~Bernard} \affiliation{University of California Berkeley, Department of Physics, Berkeley, CA 94720, USA}  
\author{A.~Bernstein} \affiliation{Lawrence Livermore National Laboratory, 7000 East Ave., Livermore, CA 94551, USA}  
\author{T.P.~Biesiadzinski} \affiliation{SLAC National Accelerator Laboratory, 2575 Sand Hill Road, Menlo Park, CA 94205, USA} \affiliation{Kavli Institute for Particle Astrophysics and Cosmology, Stanford University, 452 Lomita Mall, Stanford, CA 94309, USA} 
\author{E.M.~Boulton} \affiliation{University of California Berkeley, Department of Physics, Berkeley, CA 94720, USA} \affiliation{Lawrence Berkeley National Laboratory, 1 Cyclotron Rd., Berkeley, CA 94720, USA} \affiliation{Yale University, Department of Physics, 217 Prospect St., New Haven, CT 06511, USA}
\author{B.~Boxer} \affiliation{University of Liverpool, Department of Physics, Liverpool L69 7ZE, UK}  
\author{P.~Br\'as} \affiliation{LIP-Coimbra, Department of Physics, University of Coimbra, Rua Larga, 3004-516 Coimbra, Portugal}  
\author{S.~Burdin} \affiliation{University of Liverpool, Department of Physics, Liverpool L69 7ZE, UK}  
\author{D.~Byram} \affiliation{University of South Dakota, Department of Physics, 414E Clark St., Vermillion, SD 57069, USA} \affiliation{South Dakota Science and Technology Authority, Sanford Underground Research Facility, Lead, SD 57754, USA} 
\author{M.C.~Carmona-Benitez} \affiliation{Pennsylvania State University, Department of Physics, 104 Davey Lab, University Park, PA  16802-6300, USA}  
\author{C.~Chan} \affiliation{Brown University, Department of Physics, 182 Hope St., Providence, RI 02912, USA}  
\author{J.E.~Cutter} \affiliation{University of California Davis, Department of Physics, One Shields Ave., Davis, CA 95616, USA}  
\author{L.~de\,Viveiros}  \affiliation{Pennsylvania State University, Department of Physics, 104 Davey Lab, University Park, PA  16802-6300, USA}  
\author{E.~Druszkiewicz} \affiliation{University of Rochester, Department of Physics and Astronomy, Rochester, NY 14627, USA}  
\author{A.~Fan} \affiliation{SLAC National Accelerator Laboratory, 2575 Sand Hill Road, Menlo Park, CA 94205, USA} \affiliation{Kavli Institute for Particle Astrophysics and Cosmology, Stanford University, 452 Lomita Mall, Stanford, CA 94309, USA} 
\author{S.~Fiorucci} \affiliation{Lawrence Berkeley National Laboratory, 1 Cyclotron Rd., Berkeley, CA 94720, USA} \affiliation{Brown University, Department of Physics, 182 Hope St., Providence, RI 02912, USA} 
\author{R.J.~Gaitskell} \affiliation{Brown University, Department of Physics, 182 Hope St., Providence, RI 02912, USA}  
\author{C.~Ghag} \affiliation{Department of Physics and Astronomy, University College London, Gower Street, London WC1E 6BT, United Kingdom}  
\author{M.G.D.~Gilchriese} \affiliation{Lawrence Berkeley National Laboratory, 1 Cyclotron Rd., Berkeley, CA 94720, USA}  
\author{C.~Gwilliam} \affiliation{University of Liverpool, Department of Physics, Liverpool L69 7ZE, UK}  
\author{C.R.~Hall} \affiliation{University of Maryland, Department of Physics, College Park, MD 20742, USA}  
\author{S.J.~Haselschwardt} \affiliation{University of California Santa Barbara, Department of Physics, Santa Barbara, CA 93106, USA}  
\author{S.A.~Hertel} \affiliation{University of Massachusetts, Amherst Center for Fundamental Interactions and Department of Physics, Amherst, MA 01003-9337 USA} \affiliation{Lawrence Berkeley National Laboratory, 1 Cyclotron Rd., Berkeley, CA 94720, USA} 
\author{D.P.~Hogan} \affiliation{University of California Berkeley, Department of Physics, Berkeley, CA 94720, USA}  
\author{M.~Horn} \affiliation{South Dakota Science and Technology Authority, Sanford Underground Research Facility, Lead, SD 57754, USA} \affiliation{University of California Berkeley, Department of Physics, Berkeley, CA 94720, USA} 
\author{D.Q.~Huang} \affiliation{Brown University, Department of Physics, 182 Hope St., Providence, RI 02912, USA}  
\author{C.M.~Ignarra} \affiliation{SLAC National Accelerator Laboratory, 2575 Sand Hill Road, Menlo Park, CA 94205, USA} \affiliation{Kavli Institute for Particle Astrophysics and Cosmology, Stanford University, 452 Lomita Mall, Stanford, CA 94309, USA} 
\author{R.G.~Jacobsen} \affiliation{University of California Berkeley, Department of Physics, Berkeley, CA 94720, USA}  
\author{O.~Jahangir} \affiliation{Department of Physics and Astronomy, University College London, Gower Street, London WC1E 6BT, United Kingdom}  
\author{W.~Ji} \affiliation{SLAC National Accelerator Laboratory, 2575 Sand Hill Road, Menlo Park, CA 94205, USA} \affiliation{Kavli Institute for Particle Astrophysics and Cosmology, Stanford University, 452 Lomita Mall, Stanford, CA 94309, USA} 
\author{K.~Kamdin} \affiliation{University of California Berkeley, Department of Physics, Berkeley, CA 94720, USA} \affiliation{Lawrence Berkeley National Laboratory, 1 Cyclotron Rd., Berkeley, CA 94720, USA} 
\author{K.~Kazkaz} \affiliation{Lawrence Livermore National Laboratory, 7000 East Ave., Livermore, CA 94551, USA}  
\author{D.~Khaitan} \affiliation{University of Rochester, Department of Physics and Astronomy, Rochester, NY 14627, USA}  
\author{E.V.~Korolkova} \affiliation{University of Sheffield, Department of Physics and Astronomy, Sheffield, S3 7RH, United Kingdom}  
\author{S.~Kravitz} \affiliation{Lawrence Berkeley National Laboratory, 1 Cyclotron Rd., Berkeley, CA 94720, USA}  
\author{V.A.~Kudryavtsev} \affiliation{University of Sheffield, Department of Physics and Astronomy, Sheffield, S3 7RH, United Kingdom}  
\author{N.A.~Larsen} \email{nicole.larsen@fivesigma.net}\affiliation{Yale University, Department of Physics, 217 Prospect St., New Haven, CT 06511, USA}  \affiliation{Kavli Institute for Cosmological Physics, University of Chicago, 5620 S.~Ellis Ave, Chicago, IL 60637, USA}
\author{E.~Leason} \affiliation{SUPA, School of Physics and Astronomy, University of Edinburgh, Edinburgh EH9 3FD, United Kingdom}  
\author{B.G.~Lenardo} \affiliation{University of California Davis, Department of Physics, One Shields Ave., Davis, CA 95616, USA} \affiliation{Lawrence Livermore National Laboratory, 7000 East Ave., Livermore, CA 94551, USA} 
\author{K.T.~Lesko} \affiliation{Lawrence Berkeley National Laboratory, 1 Cyclotron Rd., Berkeley, CA 94720, USA}  
\author{J.~Liao} \affiliation{Brown University, Department of Physics, 182 Hope St., Providence, RI 02912, USA}  
\author{J.~Lin} \affiliation{University of California Berkeley, Department of Physics, Berkeley, CA 94720, USA}  
\author{A.~Lindote} \affiliation{LIP-Coimbra, Department of Physics, University of Coimbra, Rua Larga, 3004-516 Coimbra, Portugal}  
\author{M.I.~Lopes} \affiliation{LIP-Coimbra, Department of Physics, University of Coimbra, Rua Larga, 3004-516 Coimbra, Portugal}  
\author{A.~Manalaysay} \email{aaronm@lbl.gov}\affiliation{Lawrence Berkeley National Laboratory, 1 Cyclotron Rd., Berkeley, CA 94720, USA} \affiliation{University of California Davis, Department of Physics, One Shields Ave., Davis, CA 95616, USA} 
\author{R.L.~Mannino} \affiliation{Texas A \& M University, Department of Physics, College Station, TX 77843, USA} \affiliation{University of Wisconsin-Madison, Department of Physics, 1150 University Ave., Madison, WI 53706, USA} 
\author{N.~Marangou} \affiliation{Imperial College London, High Energy Physics, Blackett Laboratory, London SW7 2BZ, United Kingdom}  
\author{D.N.~McKinsey} \affiliation{University of California Berkeley, Department of Physics, Berkeley, CA 94720, USA} \affiliation{Lawrence Berkeley National Laboratory, 1 Cyclotron Rd., Berkeley, CA 94720, USA} 
\author{D.-M.~Mei} \affiliation{University of South Dakota, Department of Physics, 414E Clark St., Vermillion, SD 57069, USA}  
\author{M.~Moongweluwan} \affiliation{University of Rochester, Department of Physics and Astronomy, Rochester, NY 14627, USA}  
\author{J.A.~Morad} \affiliation{University of California Davis, Department of Physics, One Shields Ave., Davis, CA 95616, USA}  
\author{A.St.J.~Murphy} \affiliation{SUPA, School of Physics and Astronomy, University of Edinburgh, Edinburgh EH9 3FD, United Kingdom}  
\author{A.~Naylor} \affiliation{University of Sheffield, Department of Physics and Astronomy, Sheffield, S3 7RH, United Kingdom}  
\author{C.~Nehrkorn} \affiliation{University of California Santa Barbara, Department of Physics, Santa Barbara, CA 93106, USA}  
\author{H.N.~Nelson} \affiliation{University of California Santa Barbara, Department of Physics, Santa Barbara, CA 93106, USA}  
\author{F.~Neves} \affiliation{LIP-Coimbra, Department of Physics, University of Coimbra, Rua Larga, 3004-516 Coimbra, Portugal}  
\author{A.~Nilima} \affiliation{SUPA, School of Physics and Astronomy, University of Edinburgh, Edinburgh EH9 3FD, United Kingdom}  
\author{K.C.~Oliver-Mallory} \affiliation{University of California Berkeley, Department of Physics, Berkeley, CA 94720, USA} \affiliation{Lawrence Berkeley National Laboratory, 1 Cyclotron Rd., Berkeley, CA 94720, USA} 
\author{K.J.~Palladino} \affiliation{University of Wisconsin-Madison, Department of Physics, 1150 University Ave., Madison, WI 53706, USA}  
\author{E.K.~Pease} \affiliation{University of California Berkeley, Department of Physics, Berkeley, CA 94720, USA} \affiliation{Lawrence Berkeley National Laboratory, 1 Cyclotron Rd., Berkeley, CA 94720, USA} 
\author{Q.~Riffard} \affiliation{University of California Berkeley, Department of Physics, Berkeley, CA 94720, USA} \affiliation{Lawrence Berkeley National Laboratory, 1 Cyclotron Rd., Berkeley, CA 94720, USA} 
\author{G.R.C.~Rischbieter} \affiliation{University at Albany, State University of New York, Department of Physics, 1400 Washington Ave., Albany, NY 12222, USA}  
\author{C.~Rhyne} \affiliation{Brown University, Department of Physics, 182 Hope St., Providence, RI 02912, USA}  
\author{P.~Rossiter} \affiliation{University of Sheffield, Department of Physics and Astronomy, Sheffield, S3 7RH, United Kingdom}  
\author{S.~Shaw} \affiliation{University of California Santa Barbara, Department of Physics, Santa Barbara, CA 93106, USA} \affiliation{Department of Physics and Astronomy, University College London, Gower Street, London WC1E 6BT, United Kingdom} 
\author{T.A.~Shutt} \affiliation{SLAC National Accelerator Laboratory, 2575 Sand Hill Road, Menlo Park, CA 94205, USA} \affiliation{Kavli Institute for Particle Astrophysics and Cosmology, Stanford University, 452 Lomita Mall, Stanford, CA 94309, USA} 
\author{C.~Silva} \affiliation{LIP-Coimbra, Department of Physics, University of Coimbra, Rua Larga, 3004-516 Coimbra, Portugal}  
\author{M.~Solmaz} \affiliation{University of California Santa Barbara, Department of Physics, Santa Barbara, CA 93106, USA}  
\author{V.N.~Solovov} \affiliation{LIP-Coimbra, Department of Physics, University of Coimbra, Rua Larga, 3004-516 Coimbra, Portugal}  
\author{P.~Sorensen} \affiliation{Lawrence Berkeley National Laboratory, 1 Cyclotron Rd., Berkeley, CA 94720, USA}  
\author{T.J.~Sumner} \affiliation{Imperial College London, High Energy Physics, Blackett Laboratory, London SW7 2BZ, United Kingdom}  
\author{M.~Szydagis} \affiliation{University at Albany, State University of New York, Department of Physics, 1400 Washington Ave., Albany, NY 12222, USA}  
\author{D.J.~Taylor} \affiliation{South Dakota Science and Technology Authority, Sanford Underground Research Facility, Lead, SD 57754, USA}  
\author{R.~Taylor} \affiliation{Imperial College London, High Energy Physics, Blackett Laboratory, London SW7 2BZ, United Kingdom}  
\author{W.C.~Taylor} \affiliation{Brown University, Department of Physics, 182 Hope St., Providence, RI 02912, USA}  
\author{B.P.~Tennyson} \affiliation{Yale University, Department of Physics, 217 Prospect St., New Haven, CT 06511, USA}  
\author{P.A.~Terman} \affiliation{Texas A \& M University, Department of Physics, College Station, TX 77843, USA}  
\author{D.R.~Tiedt} \affiliation{University of Maryland, Department of Physics, College Park, MD 20742, USA}  
\author{W.H.~To} \affiliation{California State University Stanislaus, Department of Physics, 1 University Circle, Turlock, CA 95382, USA}  
\author{L.~Tvrznikova} \affiliation{University of California Berkeley, Department of Physics, Berkeley, CA 94720, USA} \affiliation{Lawrence Berkeley National Laboratory, 1 Cyclotron Rd., Berkeley, CA 94720, USA} \affiliation{Yale University, Department of Physics, 217 Prospect St., New Haven, CT 06511, USA}
\author{U.~Utku} \affiliation{Department of Physics and Astronomy, University College London, Gower Street, London WC1E 6BT, United Kingdom}  
\author{S.~Uvarov} \affiliation{University of California Davis, Department of Physics, One Shields Ave., Davis, CA 95616, USA}  
\author{A.~Vacheret} \affiliation{Imperial College London, High Energy Physics, Blackett Laboratory, London SW7 2BZ, United Kingdom}  
\author{V.~Velan} \affiliation{University of California Berkeley, Department of Physics, Berkeley, CA 94720, USA}  
\author{R.C.~Webb} \affiliation{Texas A \& M University, Department of Physics, College Station, TX 77843, USA}  
\author{J.T.~White} \affiliation{Texas A \& M University, Department of Physics, College Station, TX 77843, USA}  
\author{T.J.~Whitis} \affiliation{SLAC National Accelerator Laboratory, 2575 Sand Hill Road, Menlo Park, CA 94205, USA} \affiliation{Kavli Institute for Particle Astrophysics and Cosmology, Stanford University, 452 Lomita Mall, Stanford, CA 94309, USA} 
\author{M.S.~Witherell} \affiliation{Lawrence Berkeley National Laboratory, 1 Cyclotron Rd., Berkeley, CA 94720, USA}  
\author{F.L.H.~Wolfs} \affiliation{University of Rochester, Department of Physics and Astronomy, Rochester, NY 14627, USA}  
\author{D.~Woodward} \affiliation{Pennsylvania State University, Department of Physics, 104 Davey Lab, University Park, PA  16802-6300, USA}  
\author{J.~Xu} \affiliation{Lawrence Livermore National Laboratory, 7000 East Ave., Livermore, CA 94551, USA}  
\author{C.~Zhang} \affiliation{University of South Dakota, Department of Physics, 414E Clark St., Vermillion, SD 57069, USA}